\documentclass[article]{jss}
\usepackage[utf8]{inputenc}

\providecommand{\tightlist}{%
  \setlength{\itemsep}{0pt}\setlength{\parskip}{0pt}}

\author{
Kristian Brock\\Cancer Research UK Clinical Trials Unit, University of Birmingham
}
\title{\pkg{trialr}: Bayesian Clinical Trial Designs in R and Stan}

\Plainauthor{Kristian Brock}
\Plaintitle{trialr: Bayesian Clinical Trial Designs in R and Stan}

\Abstract{
This manuscript introduces an \proglang{R} package called \pkg{trialr}
that implements a collection of clinical trial methods in
\proglang{Stan} and \proglang{R}. In this article, we explore three
methods in detail. The first is the continual reassessment method for
conducting phase I dose-finding trials that seek a maximum tolerable
dose. The second is EffTox, a dose-finding design that scrutinises doses
by joint efficacy and toxicity outcomes. The third is the augmented
binary method for modelling the probability of treatment success in
phase II oncology trials with reference to repeated measures of
continuous tumour size and binary indicators of treatment failure. We
emphasise in this article the benefits that stem from having access to
posterior samples, including flexible inference and powerful
visualisation. We hope that this package encourages the use of Bayesian
methods in clinical trials.
}

\Plainkeywords{clinical trial, bayesian, dose finding, phase II, R, Stan}

\Address{
    Kristian Brock\\
  Cancer Research UK Clinical Trials Unit, University of Birmingham\\
  Birmingham, B15 2TT, UK\\
  E-mail: \email{k.brock@bham.ac.uk}\\
  URL: \url{https://github.com/brockk/}\\~\\
  }

\usepackage{amsmath}

\begin{document}

\hypertarget{introduction}{%
\section{Introduction}\label{introduction}}

Clinical trials are sequential medical experiments in humans. Their
collective goal is to identify experimental therapies that offer
sufficient benefit to warrant use in standard clinical practice.
Fundamental to this is the estimation of treatment efficacy and safety.
Many experimental designs have been proposed to these ends.

\proglang{Stan} \citep{Carpenter2016} is a probabilistic programming
language allowing full Bayesian inference via Markov chain Monte Carlo
(MCMC) sampling. We present in this article the \pkg{trialr} package
implementing several clinical trial designs in \proglang{R} and
\proglang{Stan}.

Investigation of safety is the major preoccupation of early phase
clinical trials. A typical phase I trial would conduct a dose-finding
experiment to identify a dose of a novel therapy that is associated with
an acceptable adverse event (AE) rate. The term \textit{dose} here is
used loosely to reflect the general intensity of treatment. In most
cases, dose would have the intuitive property of reflecting the quantity
of a molecule that is administered to the patient in a treatment window.
It could, however, represent the frequency or the duration of
administration, or a combination of the two. It could also contain
information on those aspects of several experimental therapies that are
given together.

In many instances, a dose-finding experiment would be designed to
incorporate dose escalation, where successively higher doses are
sequentially investigated in cohorts of patients. Such experimental
designs typically assess the patient-level presence or absence of
\textit{dose-limiting toxicity} (DLT), the manifestation of at least one
of potentially many prespecified adverse events that would be considered
sufficiently serious to halt treatment administration. For illustration,
treatment could commence at a low dose for which there is reasonable
expectation of safety. If a sufficiently low number of these patients
experience DLT, the dose may be escalated for the next group of
patients. Dose-finding trials typically continue sequentially, with
successive doses being chosen in response to the outcomes observed thus
far. As such, dose-finding trials are one of the more simple types of
\textit{adaptive} trial, where some treatment parameter for patients
(here, dose) is affected by the outcomes observed in earlier patients.

A key assumption of dose escalation and de-escalation solely with
reference to DLT is that a higher dose is more likely to provide
clinical benefit. A higher dose is accepted as being more attractive
than a lower dose so long as both are associated with an acceptable rate
of DLT, without any reference to the level of efficacy expected or
observed. Logic, and the toxicologists' adage
\textit{sola dosis facit venenum} or \textit{the dose makes the poison},
dictate that the probability of toxicity increases in dose. These
so-called \textit{monotonicity assumptions} are the cornerstones of
traditional phase I dose-finding. The highest dose with predicted DLT
probability less than or close to some target rate is referred to as the
maximum tolerable dose (MTD).

Identification of the MTD is a typical objective of phase I trials.
There are several experimental designs proposed to achieve this goal.
The most common is 3+3 \citep{carterStudyDesignPrinciples1973}, a simple
algorithm that provides rules for changing dose based on the DLT
outcomes of cohorts of 3 patients. It has been widely criticised for its
relatively poor operating performance, its non-statistical nature, and
its lack of flexibility
\citep{OQuigley1990, oquigleyExperimentalDesignsPhase2006, Iasonos2008, LeTourneau2009}.
Despite this, it stubbornly persists as the most frequently used
dose-finding design \citep{Rogatko2007, Chiuzan2017}, due in no small
part to its simplicity and familiarity.

There are also many statistical designs that seek the MTD
\citep[amongst others]{OQuigley1990, Tighiouart2010, Ji2013a}. The
Continual Reassessment Method (CRM) by \cite{OQuigley1990} is
particularly noteworthy because of the great amount of innovation it has
facilitated, including the time-to-event CRM (TITE-CRM) approach
\citep{Cheung2000} for studying late-onset toxic events.

Modern cancer drugs like targeted therapies and immunotherapies do not
necessarily support the monotonic efficacy assumption. For instance, two
doses of the immunotherapy drug pembrolizumab that differed by a factor
of five were shown in a large randomised trial to have very similar
response rates \citep{Herbst2016}. When conducting dose-finding in
therapies like these, there may be motivation to scrutinise doses by
efficacy and toxicity outcomes. There are several designs that perform
this task \citep{Braun2002, Thall2004, Zhang2006, Wages2014}.

Traditional phase II trials take the dose recommended in phase I and
investigate early signs of treatment activity or efficacy. The objective
here is to screen out treatments that are obviously ineffective. These
trials can generally be single-arm or randomised. The onus is on
conducting the trial and disseminating the results quickly so that
onward clinical research can proceed promptly. As such, sample sizes
tend to be small and outcomes that can be assessed over the short-term
are generally preferred.

Finally, phase III trials seek to provide evidence good enough to affect
medical practice. Most commonly, phase III trials test whether an
experimental treatment is superior to the current standard of care.
These trials commonly randomise patients to one of several treatment
arms, use clinically-relevant long-term outcome measures, and have large
sample sizes.

Bayesian methods in clinical trials are used predominantly, although not
exclusively, in phases I and II. This is due in part to the limited
amount of information available from the low sample sizes used. Using
prior information to complement an analysis can be desirable, and the
model-based methods can benefit computationally from the Bayesian
approach.

In this article, we focus on three clinical trial models implemented in
\pkg{trialr} under the Bayesian paradigm using \proglang{R} and
\proglang{Stan}:

\begin{itemize}
\tightlist
\item
  the CRM and TITE-CRM models for traditional toxicity-oriented
  dose-finding;
\item
  the EffTox model for efficacy and toxicity dose-finding;
\item
  the Augmented Binary (AugBin) method \citep{Wason2013} for analysing
  dichotomous response in phase II cancer trials whilst retaining the
  continuous information in measurements of tumour size.
\end{itemize}

There are several other R-packages that implement the Bayesian CRM in
\proglang{R}, including \pkg{dfcrm} \citep{dfcrm, Cheung2011},
\pkg{bcrm} \citep{bcrm}, and \pkg{crmPack} \citep{crmPack}. \pkg{dfcrm}
uses a non-MCMC approach. \pkg{bcrm} and \pkg{crmPack} support MCMC via
\proglang{JAGS} and \proglang{WinBUGS}. To the best of our knowledge,
\pkg{trialr} is the only package on CRAN to implement the CRM and EffTox
in \proglang{Stan}, and the only package on CRAN to implement the AugBin
method.

To avoid large-scale repetition of material in previous publications on
CRM software \citep{bcrm, Cheung2011}, we focus the dose-finding aspects
of this article on functions related to dose transtion pathways
\citep{Yap2017, Brock2017a}, a tool to aid trial planning. This approach
is considered for both the CRM and EffTox designs. Then we introduce the
AugBin method for single arm trials with two post-baseline analyses.

\pkg{trialr} was first added to CRAN in 2017 with Stan implementations
of EffTox, a hierarchical model for analysing responses to a single drug
in several related subtypes of a disease \citep{Thall2003}, and a
joint-model for analysing co-primary efficacy and toxicity outcomes when
baseline covariate information is available
\citep{brockMethodsIncreaseEfficiency}. Successive updates have seen the
addition of the CRM and AugBin models. \pkg{trialr} relies heavily on
\pkg{tidyverse} \citep{tidyverse} packages to create a modern workflow
in \proglang{R} for Bayesian approaches to clinical trials.

\newpage

\hypertarget{methodology}{%
\section{Methodology}\label{methodology}}

\hypertarget{crm}{%
\subsection{CRM}\label{crm}}

The Continual Reassessment Method (CRM) was originally published by
\cite{OQuigley1990} to conduct dose-finding trials seeking a maximum
tolerable dose (MTD). It is a truly seminal design, with many variants
appearing over the years to handle different clinical scenarios, such as
related groups \citep{oquigleyContinualReassessmentMethod2003},
late-onset toxicity \citep{Cheung2000}, efficacy and toxicity outcomes
\citep{Braun2002, Zhang2006}, and more. In this section, we focus on the
original Bayesian design that seeks the MTD in an homogeneous patient
group.

In a trial of doses \(d_1, ..., d_K\), the investigators provide their
best guess of the probability of DLT at each dose to create a toxicity
\textit{skeleton} \((p_1, ..., p_K)\). CRM assumes monotonically
increasing toxicity risk so we stipulate that \(p_1 < ... < p_K\).

The probability of DLT at dose \(x\) is modelled to be \(F(x, \theta)\),
where \(F\) is a smooth mathematical function, and \(\theta\) is a
general vector of parameters. Different variants of this model use
different forms for \(F\), and different prior distributions on
\(\theta\). The CRM variants currently implemented in \texttt{trialr}
are listed in Table \ref{tab:crm}.

\begin{table}
  \begin{center}
    \begin{tabular}{| l | l | l |}
      \hline
      \texttt{model} & Link function & Parameters \\ 
      \hline
      \texttt{empiric} & $F(d, \beta)$ = $d^{\exp{\beta}}$ & normal prior on $\beta$ \\  
      \texttt{logistic} & $F(d, \beta) = 1 / (1 + \exp{(-a_0 - \exp{(\beta)} d}))$ & $a_0$ fixed, normal prior on $\beta$ \\
\texttt{logistic\_gamma} & $F(d, \beta) = 1 / (1 + \exp{(-a_0 - \beta d)})$ & $a_0$ fixed, gamma prior on $\beta$ \\ 
\texttt{logistic2} & $F(d, \alpha, \beta) = 1 / (1 + \exp{(-\alpha - \exp{(\beta)} d)})$ & normal priors on $\alpha$ and $\beta$ \\ 
      \hline
    \end{tabular}
  \end{center}
  \caption{CRM models implemented in \pkg{trialr}.}
  \label{tab:crm}
\end{table}

To ensure that the probability of toxicity increases with dose, the
slope parameter \(\beta\) is exponentiated in some of the models in
Table \ref{tab:crm}. In the model with gamma prior, this step is not
necessary because the prior constrains the posterior to admit only
positive values.

For the purposes of statistical modelling, the explanatory dose
variables \(d_k\) are codified such that

\[ p_k = F(d_k, \hat{\theta}) , \]

where \(\hat{\theta}\) is the vector of prior expected values of the
elements of \(\theta\). That is, the \(d_k\) are derived so that using
the prior expected value of each model parameter, the prior expected
probability of DLT at \(d_k\) is \(p_k\). \citet{Cheung2011} calls the
\(d_k\) \textit{dose-labels} to distinguish them from the actual dose
units like ``10 mg'' that a clinician might actually administer to a
patient.

For patient \(i\), let \(Y_i\) be a random variable taking values
\(\{0, 1\}\) reflecting the absence and presence of DLT, and
\(x_i \in \left\{ d_1, ..., d_K \right\}\) be the dose administered.
After the evaluation of \(I\) patients, the aggregate likelihood is:

\[ L_I(\theta) = \prod_{i=1}^I \left\{ F(x_i, \theta) \right\}^{Y_i} \left\{ 1 - F(x_i, \theta) \right\}^{1-Y_i} \]

This quantity can either be maximised with respect to \(\theta\) to
obtain the maximum likelihood estimate, or augmented with priors and
analysed in a Bayesian way. CRM was originally published as a Bayesian
design and that is the approach we take in \texttt{trialr} using Stan.

\citet{Cheung2000} introduced the notion of weighting the \(Y_i\) to
reflect the extent to which they have been evaluated. For instance, if a
patient is 50\% of the way through the evaluation period without having
yet experienced toxicity, we might interpret them as providing an
observation \(Y_i = 0\) with weight 0.5. This leads to the weighted
likelihood statement:
\[ L_I(\theta) = \prod_{i=1}^I \left\{ w_i F(x_i, \theta) \right\}^{Y_i} \left\{ 1 - w_i F(x_i, \theta) \right\}^{1-Y_i}, \]
as used in the TITE-CRM design. This insight faciltated the study of
long-term toxicity events, without the requirement that each patient
complete the evaluation period before they contribute information to the
dose selection decision.

CRM models are accessed in \pkg{trialr} primarily through the
\texttt{stan\_crm} function, as described below.

\hypertarget{efftox}{%
\subsection{EffTox}\label{efftox}}

The EffTox design \citep{Thall2004, Thall2006, Thall2014} conducts
dose-finding clinical trials where both efficacy and toxicity events
guide dose selection. This contrasts methods like 3+3 and CRM where dose
selection is determined by toxicity events only. It is no longer assumed
that higher doses are implicitly more likely to be beneficial to
patients. Instead, EffTox calculates attractiveness or \textit{utility}
scores for doses that trade-off the estimated probabilities of efficacy
and toxicity. The design iteratively seeks the dose with the highest
utility score. We will call this the \textit{optimum biological dose}
(OBD) to distinguish it from the MTD and the pratice of escalating under
assumed montonically increasing efficacy. When recommending the next
dose, EffTox does not blindly recommend the prevailing OBD. Instead, it
uses rules that forbid the skipping of untested doses. Full details are
given in the authors' publications but we provide a brief recap of the
statistical methodology here.

For doses \((y_1, ..., y_n)\), the authors define standardised doses
\((x_1, ..., x_n)\) using the transform

\[x_j = \log{y_j} - \sum_{k=1}^n \frac{\log{y_k}}{n}\]

The model uses six parameters,
\(\theta = (\alpha, \beta, \gamma, \zeta, \eta, \epsilon)\). At dose
\(x\), the marginal probabilities of toxicity and efficacy are estimated
by the logit models:

\[\text{logit } \pi_T(x, \theta) = \alpha + \beta x\] and

\[\text{logit } \pi_E(x, \theta) = \gamma + \zeta x + \eta x^2\]

Let \(\boldsymbol{Y} = (Y_E, Y_T)\) be indicators of binary efficacy and
toxicity events and let
\(\pi_{a, b}(x, \theta) = \text{Pr}(Y_E = a, Y_T = b | x, \theta)\) for
\(a, b \in \left\{ 0, 1\right\}\). The efficacy and toxicity events are
associated by the joint probability function

\[\pi_{a,b}(x, \theta) = (\pi_E)^a (1-\pi_E)^{1-a} (\pi_T)^b (1-\pi_T)^{1-b} + \\ (-1)^{a+b} (\pi_E) (1-\pi_E) (\pi_T) (1-\pi_T) \frac{e^\psi-1}{e^\psi+1}\],

where \(\psi\) is a parameter that measures the strength of association
between the co-primary outcomes, and \((x, \theta)\)-notation has been
suppressed on the right-hand side for brevity. Normal priors are
specified for the elements of \(\theta\).

Let \(\mathcal{D}\) represent the trial data collected hitherto. At each
dose update decision, the dose \(x\) is acceptable if

\[\text{Pr}\left\{ \pi_T(x, \boldsymbol{\theta}) < \overline{\pi}_T | \mathcal{D} \right\} > p_T,\]

\[\text{Pr}\left\{ \pi_E(x, \boldsymbol{\theta}) > \underline{\pi}_E | \mathcal{D} \right\} > p_E\]

and is no more than than one position below the lowest dose-level given
and no more than one position above the highest dose-level given. The
effect of these last two criteria is that untried doses may not be
skipped in escalation or de-escalation. The values \(\underline{\pi}_E\)
and \(\overline{\pi}_T\) are provided by the investigators with
reference to what is acceptable in the clinical scenario, given the set
of alternative treatments. \citet{Thall2004} selected
\(p_E = p_T = 0.1\) so that doses were excluded from the acceptable set
only if we are very sure that they are either inactive or excessively
toxic.

The utility of dose \(x\), with efficacy \(\pi_E(x, \theta)\) and
toxicity \(\pi_T(x, \theta)\) is
\[u(\pi_E, \pi_T) = 1 - \left( \left(\frac{1-\pi_E}{1-\pi_{1,E}^*}\right)^p + \left(\frac{\pi_T}{\pi_{2,T}^*}\right)^p \right) ^ \frac{1}{p}\]

where \(p\) is calculated to intersect the points \((\pi_{1,E}^*, 0)\),
\((1, \pi_{2,T}^*)\) and \((\pi_{3,E}^*, \pi_{3,T}^*)\) in the
efficacy-toxicity plane. We will refer to these as \emph{hinge points}
but that is not nomenclature used by the authors.

At the dose selection decision, the dose-level from the subset of doses
that are acceptable with maximal utility is selected to be given to the
next patient or cohort. If there are no acceptable doses, the trial
stops and no dose is recommended.

There are several published EffTox examples, including explanations and
advice on parameter choices
\citep{Thall2004, Thall2006, Thall2014, Brock2017a}.

The MD Anderson Cancer Center publishes
software\footnote{see https://biostatistics.mdanderson.org/SoftwareDownload/SingleSoftware/Index/2}
to perform calculations and simulations for EffTox trials. The software
is available for Windows in compiled-form. Thus, trialists cannot run
the software on Mac or Linux unless via a virtual machine or emulator.
Furthermore, they may not alter the behaviour of the model.
\citet{Brock2017a} describe a clinical trial scenario where some
alteration to the default model behaviour would have been preferable. It
was this that prompted the author to write the open-source
implementation provided in \texttt{trialr}.

\hypertarget{augmented-binary-method}{%
\subsection{Augmented Binary method}\label{augmented-binary-method}}

In oncology clinical trials, tumours are often measured at baseline and
repeatedly post-baseline to gauge the response to therapy. The Response
Evaluation Criteria in Solid Tumors (RECIST)\citep{Eisenhauer2009}
define an algorithm for classifying into categories the tumour
measurements compared to baseline. Up to five so-called
\textit{target lesions} are identified at baseline and the sum of their
largest diameters is recorded. A \textit{complete response} is said to
have occurred post-baseline when all tumour lesions disappear. A
\textit{partial response} occurs when the aggregate diameter of the
target lesions shrinks by at least 30\% and no new lesions are detected.
Commonly, the complete and partial response events are combined to form
an \textit{objective response} category, and this is used to calculate
the binary response rate in a trial.

Statisticians are aware that this dichotomisation of a continuous
underlying measure leads to information loss and less efficient
analysis. However, responder analysis is endemic in cancer trials. In
response, \citet{Wason2013} introduced the Augmented Binary model. It
analyses the continuous log-tumour size ratio, whilst building in
mechanisms for so-called \textit{non-shrinkage failures} that signal
lack of success, such as the manifestation of new lesions, even if
requisite tumour shrinkage is observed. Wason \& Seaman introduced
versions of their model for single-arm and randomised groups trials. We
describe here the single-arm variant.

Let \(z_{0i}\) be the tumour size at baseline for patient \(i\),
\(z_{1i}\) the size at interim, and \(z_{2i}\) the size at the end of
the trial. Let
\((y_{1i}, y_{2i}) = (\log{(z_{1i} / z_{0i})}, \log{(z_{2i} / z_{0i})})\)
be the log tumour size ratios with respect to baseline. Furthermore, the
authors define variables \(D_{1i}\) and \(D_{2i}\) to represent the
observation of non-shrinkage failure at interim and final assessments.
Finally, they define \(S_i\) to be a composite success indicator. Using
30\% shrinkage as the threshold for success, \(S_i = 1\) if
\(D_{1i} = D_{2i} = 0\) and \(y_{2i} < \log{(0.7)}\).

The authors specify the following statistical models. The tumour
shrinkage variables are modelled using a bivariate normal distribution:

\[ (y_{1i}, y_{2i})^T \sim N\left( (\mu_{1i}, \mu_{2i})^T, \Sigma\right),\]
where \(\mu_{1i} = \alpha + \gamma z_{0i}\) and
\(\mu_{2i} = \beta + \gamma z_{0i}\), and \(\Sigma\) is assumed to be
unstructured.

The non-shrinkage failures are estimated using logit models:

\[ \text{logit} \; \text{Pr}(D_{1i} = 1 | Z_{0i}) = \alpha_{D1} + \gamma_{D1} z_{0i} \]
and

\[ \text{logit} \; \text{Pr}(D_{2i} = 1 | D_{1i} = 0, Z_{0i}, Z_{1i}) = \alpha_{D2} + \gamma_{D2} z_{1i} .\]

Finally,

\[ \text{Pr}(S_i = 1 | z_{0i}, \theta) = \int_{-\infty}^\infty \int_{-\infty}^\infty \text{Pr}(S_i = 1 | z_{0i}, y_{1i}, y_{2i}, \theta) f_{Y_1, Y_2}(y_{1i}, y_{2i}) dy_{1i} dy_{2i} .\]

In their paper, Wason \& Seaman fit the model using the \proglang{R}
functions \texttt{gls} and \texttt{glm}, and used cubature to resolve
the double integral above. In \pkg{trialr}, we provide priors on the
elements of \(\theta\) and sample from the posterior using
\proglang{Stan}.

\hypertarget{using-trialr}{%
\section{Using trialr}\label{using-trialr}}

\hypertarget{describing-outcomes-in-dose-finding-trials}{%
\subsection{Describing outcomes in dose-finding
trials}\label{describing-outcomes-in-dose-finding-trials}}

In the following sections, we will make frequent use of the following
scheme for describing outcomes in dose-finding trials. In situations
where toxicity only is being assessed, as with CRM, we will use the
characters \texttt{T} and \texttt{N} to represent the outcomes toxicity
and no-toxicity. We will string these characters behind integer
dose-levels to represent the outcomes of patients in cohorts treated at
a dose. For instance, the string \texttt{1NNN\ 2TNT} represents the
outcomes of six patients, treated in two cohorts of three. The first
cohort was treated at dose-level 1 and none of the patients had
toxicity. The second cohort received dose-level 2 and two of the three
experienced toxicity.

In the setting where efficacy and toxicity outcomes are valuated, as
with EffTox, we use the characters \texttt{E}, \texttt{T}, \texttt{B},
and \texttt{N} to represent the outcomes efficacy only, toxicity only,
both and neither. Again, we string these characters behind integer
dose-levels to represent cohorts. The string \texttt{1NN\ 2EB}
represents the outcomes of two cohorts of two patients. The second
cohort was treated at dose-level 2 and both experienced the efficacy
event, but one of the two also experienced toxicity.

This syntax was originally described in the context of an EffTox trial
\citep{Brock2017a}. It is used in \pkg{trialr} to describe outcomes in
dose-finding trials.

\hypertarget{crm-1}{%
\subsection{CRM}\label{crm-1}}

The core function for fitting a CRM model in \pkg{trialr} is
\code{stan_crm}. To demonstrate usage, we will fit a model to the
scenario described by \citet[p.~21]{Cheung2011}.

In a five-dose scenario, they seek the dose associated with the target
toxicity level 25\% using the initial toxicity skeleton (5\%, 12\%,
25\%, 40\%, 55\%). They use a one-parameter logistic model with
intercept fixed at 3 and a \(N(0, 1.34)\) prior distribution on
\(\beta\). In their scenario, they have already evaluated five patients
at dose-levels 3, 4 and 5 with outcomes \texttt{3N\ 5N\ 5T\ 3N\ 4N} and
they want to know the dose recommended by the model for the sixth
patient.

\begin{CodeChunk}

\begin{CodeInput}
R> library(trialr)
R> 
R> skeleton <- c(0.05, 0.12, 0.25, 0.40, 0.55)
R> target <- 0.25
R> fit <- stan_crm(outcome_str = '3N 5N 5T 3N 4N',
R+                 skeleton = skeleton, target = target,
R+                 model = 'logistic', a0 = 3,
R+                 beta_mean = 0, beta_sd = sqrt(1.34),
R+                 seed = 123, refresh = 0)
R> 
R> fit
\end{CodeInput}

\begin{CodeOutput}
  Patient Dose Toxicity Weight
1       1    3        0      1
2       2    5        0      1
3       3    5        1      1
4       4    3        0      1
5       5    4        0      1

  Dose Skeleton N Tox ProbTox MedianProbTox ProbMTD
1    1     0.05 0   0  0.0343       0.00833   0.043
2    2     0.12 0   0  0.0697       0.02838   0.074
3    3     0.25 2   0  0.1371       0.08599   0.161
4    4     0.40 1   0  0.2295       0.18900   0.246
5    5     0.55 2   1  0.3507       0.34000   0.476

The model targets a toxicity level of 0.25.
The dose with estimated toxicity probability closest to target is 4.
The dose most likely to be the MTD is 5.
Model entropy: 1.32
\end{CodeOutput}
\end{CodeChunk}

\pkg{trialr} currently supports models listed in Table \ref{tab:crm}.
The choice of \texttt{model} determines what model parameters are
required. We see above that the \texttt{logistic} model requires the
parameter \texttt{a0} for the fixed intercept, and \texttt{beta\_mean}
and \texttt{beta\_sd} parameters to describe the prior on \(\beta\). In
contrast, the \texttt{logistic2} parameterisation requires prior
hyperparameters on \(\alpha\) and \(\beta\). See \texttt{?\ stan\_crm}
for full details.

Extra parameters in calls to \texttt{stan\_crm} are passed onwards via
the \texttt{...} operator to the function in \pkg{rstan} that performs
the MCMC sampling, \texttt{rstan::sampling}. The parameter choice
\texttt{seed\ =\ 123} allows reproducible sampling. The effect of
\texttt{refresh\ =\ 0} is to suppress the log messages created by
\proglang{Stan}.

The object returned by \texttt{stan\_crm} has type \texttt{crm\_fit},
which subclasses the type \texttt{dose\_finding\_fit}. Its default
\texttt{print} method displays a table of patient-level information,
followed by a table of dose-level information, followed by some summary
information. \texttt{ProbTox} is the posterior mean probability of
toxicity at each dose calculated from the full posterior distributions.

\pkg{trialr} works with \pkg{tidybayes} \citep{tidybayes} so that
posterior samples are simple to access and flexible to work with.

\begin{CodeChunk}

\begin{CodeInput}
R> library(dplyr)
R> library(tidybayes)
R> fit %>% 
R+   gather_draws(prob_tox[dose]) %>% 
R+   head
\end{CodeInput}

\begin{CodeOutput}
# A tibble: 6 x 6
# Groups:   dose, .variable [5]
  .chain .iteration .draw  dose .variable   .value
   <int>      <int> <int> <int> <chr>        <dbl>
1      1          1     1     1 prob_tox  0.0103  
2      1          1     1     2 prob_tox  0.0339  
3      1          1     1     3 prob_tox  0.0986  
4      1          1     1     4 prob_tox  0.209   
5      1          1     1     5 prob_tox  0.363   
6      1          2     2     1 prob_tox  0.000737
\end{CodeOutput}
\end{CodeChunk}

This is particularly valuable because it facilitates inference and
visualisation. For instance, each posterior sample for \(\beta\)
generates a dose-DLT curve, and each of these nominates a candidate for
the MTD, being the dose with estimated Prob(DLT) closest to the toxicity
target. We can plot some of these generated curves, colouring each by
the MTD that they nominate, to visualise the level of uncertainty still
inherent in the situation at this early stage.

\begin{CodeChunk}

\begin{CodeInput}
R> fit %>% 
R+   gather_draws(prob_tox[dose]) %>% 
R+   group_by(.draw) %>% 
R+   summarise(mtd = dose[which.min(abs(.value - target))]) -> mtd_candidates
R> 
R> library(ggplot2)
R> fit %>% 
R+   gather_draws(prob_tox[dose]) %>% 
R+   left_join(mtd_candidates, by = '.draw') %>% 
R+   filter(.draw <= 200) %>% 
R+   ggplot(aes(x = dose, y = .value, group = .draw)) +
R+   geom_line(aes(col = as.factor(mtd)), alpha = 0.5) + 
R+   geom_hline(yintercept = target, col = 'red', linetype = 'dashed') + 
R+   labs(title = 'The identify of the MTD is still shrouded in mystery', 
R+        y = 'Prob(DLT)', col = 'MTD') +
R+   theme(legend.position = 'bottom')
\end{CodeInput}
\begin{figure}

{\centering \includegraphics{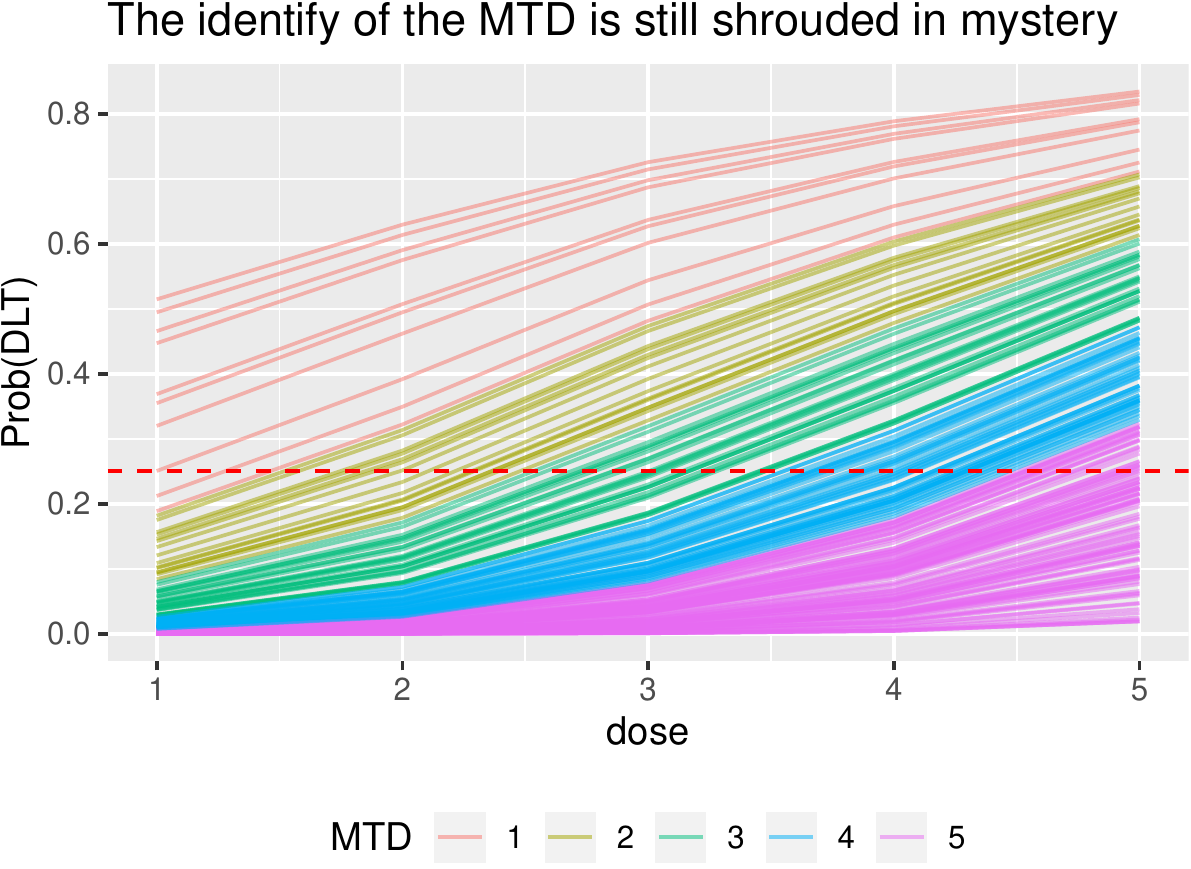} 

}

\caption[Posterior dose-toxicity curves generated by a CRM model fit to five patients' data]{Posterior dose-toxicity curves generated by a CRM model fit to five patients' data.}\label{fig:unnamed-chunk-3}
\end{figure}
\end{CodeChunk}

We plot just 200 curves to avoid saturating the plot. We see that curves
that nominate each of the five doses as the implied MTD are readily
generated.

Using each of the posterior dose-DLT curves, let us calculate the
posterior probability that each dose is the MTD according to this model.

\begin{CodeChunk}

\begin{CodeInput}
R> mtd_candidates %>% 
R+   count(mtd) %>% 
R+   mutate(prob_mtd = n / sum(n))
\end{CodeInput}

\begin{CodeOutput}
# A tibble: 5 x 3
    mtd     n prob_mtd
  <int> <int>    <dbl>
1     1   172    0.043
2     2   296    0.074
3     3   643    0.161
4     4   984    0.246
5     5  1905    0.476
\end{CodeOutput}
\end{CodeChunk}

This matches the \texttt{ProbMTD} statistics in the model output above.

\citet{Yap2017} introduced \textit{dose transition pathways} (DTPs) as
the link between model-based statistical dose-finding algorithms and
simple decision making. A pathway in a dose-finding trial is simply a
sequence of doses that were delivered to patients and the outcomes that
the patients experienced.

Consider a trial that has already treated four patients in two cohorts:

\begin{CodeChunk}

\begin{CodeInput}
R> outcomes <- '2NN 3TN'
\end{CodeInput}
\end{CodeChunk}

Beyond questioning what dose will be advised for the third cohort, we
might wonder about all the doses could possibly be advised for the
fourth and fifth cohorts. This might have material consequences for the
amount of drug that is to be manufactured and shipped to sites.
\pkg{trialr} provides flexible methods to automate such analyses.

For the purposes of this example, let us assume we are using the
\texttt{empiric} model form and a unit normal prior on \(\beta\), and
the following parameters:

\begin{CodeChunk}

\begin{CodeInput}
R> skeleton <- c(0.05, 0.15, 0.25, 0.4, 0.6)
R> target <- 0.25
\end{CodeInput}
\end{CodeChunk}

We calculate DTPs using the \texttt{crm\_dtps} method:

\begin{CodeChunk}

\begin{CodeInput}
R> paths1 <- crm_dtps(skeleton = skeleton, target = target, model = 'empiric', 
R+                    cohort_sizes = c(3, 3), previous_outcomes = outcomes,
R+                    beta_sd = 1, refresh = 0)
\end{CodeInput}
\end{CodeChunk}

This function uses many of the same parameters as \texttt{stan\_crm}.
The \texttt{previous\_outcomes} parameter describes the path the trial
has already taken; each calculated future path starts from here. It can
be omitted to reflect a trial that has not yet started. The
\texttt{cohort\_sizes} parameter controls the number and size of future
cohorts to consider. Each cohort of \(n\) may result in 0, 1,
2,\ldots{}, or \(n\) patients experiencing DLT. Thus calculating
transitions for two future cohorts of three patients, as above, yields
16 possible paths.

At the end of each cohort, there is a dose selection decision. By
default, CRM models choose the dose with posterior mean probability of
toxicity closest to the target. However, in practice, trialists may want
to tailor this decision.

Custom dose selection is supported in \texttt{crm\_dtps} via the
\texttt{user\_dose\_func} parameter. For instance, investigators may
want to prevent the skipping of untested doses in escalalation, or build
in the facility to stop a trial early if there is sufficient evidence
that a reference dose is excessively toxic. Both of these design aspects
were used in a recently published dose-escalation trial in acute myeloid
leukaemia \citep{craddockCombinationLenalidomideAzacitidine2019} and
they are provided in \texttt{trialr} by the \texttt{careful\_escalation}
function. It prevents skipping in escalation by recommending, where
appropriate, the next highest dose to the maximum that has yet been
used. For instance, if the default behaviour is to escalate from dose 2
to dose 4 without having tested dose 3, \texttt{careful\_escalation}
will instead recommend dose 3. Additionally, it returns \texttt{NA} if a
reference dose is detected to be too toxic and this is interpreted by
\texttt{crm\_dtps} as the sign that this trial path should stop. Users
are free to provide any delegate function via \texttt{user\_dose\_func}
that takes a \texttt{dose\_finding\_fit} as the sole parameter and
returns either an integer dose-level or \texttt{NA}. We re-calculate
those paths using our custom dose function:

\begin{CodeChunk}

\begin{CodeInput}
R> paths2 <- crm_dtps(skeleton = skeleton, target = target, model = 'empiric',
R+                    cohort_sizes = c(3, 3), previous_outcomes = outcomes,
R+                    user_dose_func = function(x) {
R+                      careful_escalation(x, tox_threshold = target + 0.1, 
R+                                         certainty_threshold = 0.7,
R+                                         reference_dose = 1)
R+                    }, beta_sd = 1, seed = 123, refresh = 0)
\end{CodeInput}
\end{CodeChunk}

Here, a pathway will stop if ever \begin{equation}
Pr( F(d_1, \beta) > 0.35 | \mathcal{D}) > 0.7,
\end{equation} i.e.~if there is at least a 70\% chance that the
probability of DLT at the lowest dose exceeds 35\%.

These paths are shown in Figure \ref{fig:crm_dtps_graph}. Code to
recreate this graph appears in the Appendix. We see that the subsequent
path \texttt{2TTT\ 1TTT} yields the understandable decision to stop, but
that every other path selects a dose. We will only advise as high as
dose 4 if no DLTs are seen in the next 6 patients. This level of
foresight and transparency underpins our motivation to use dose
pathways, particularly via an effective visualisation like Figure
\ref{fig:crm_dtps_graph}.

\begin{figure}
  \centering
  \includegraphics{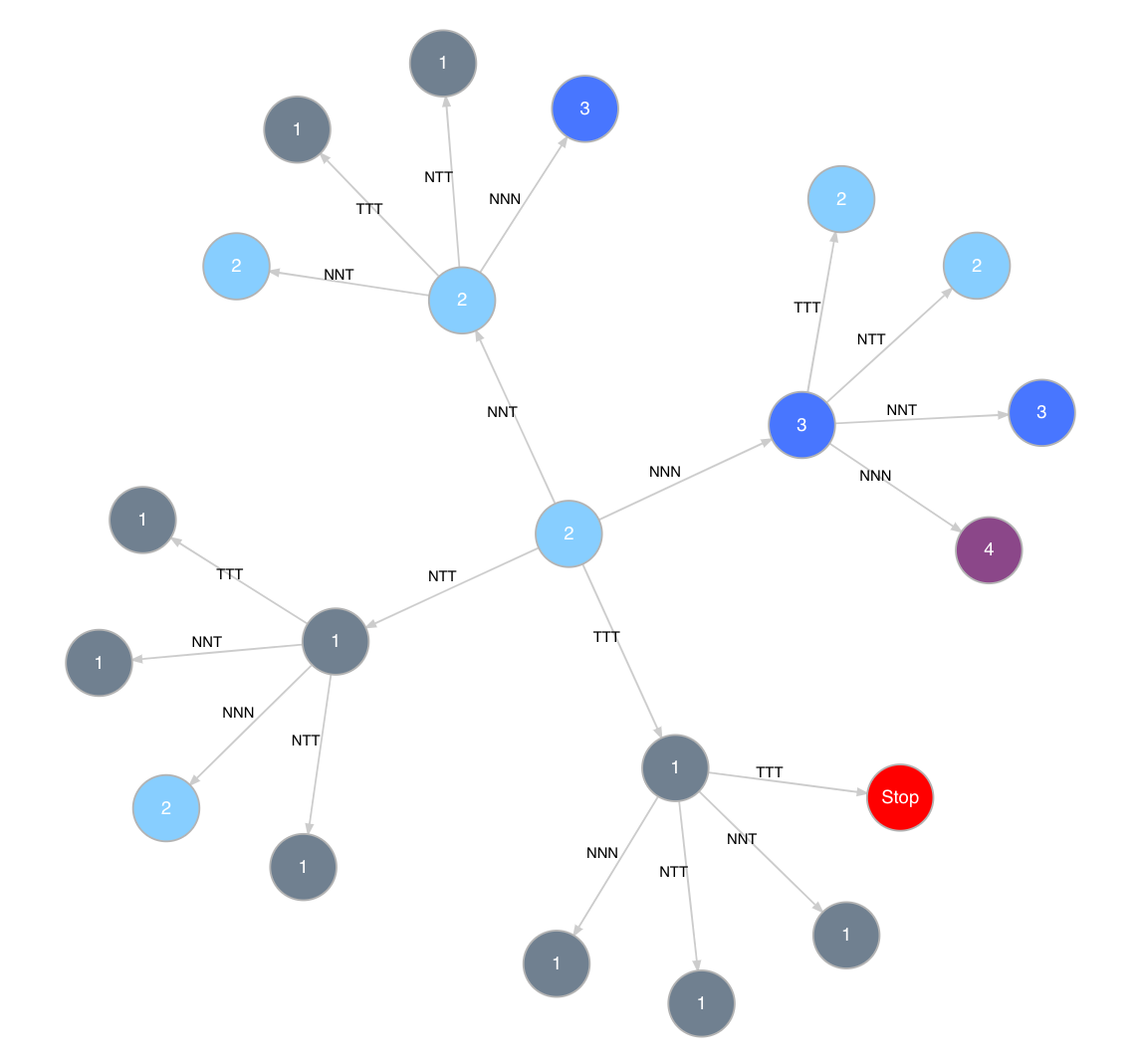}
  \caption{Dose transition pathways for the next two cohorts of three patients in a dose-finding trial using a CRM design.}
  \label{fig:crm_dtps_graph}
\end{figure}

The returned object, \texttt{paths2}, has type
\texttt{dose\_finding\_paths} and contains model fits at each node on
each pathway, as well as information on the tree structure. This is
perhaps best viewed as a \texttt{tibble}:

\begin{CodeChunk}

\begin{CodeInput}
R> library(tibble)
R> paths2_df <- as_tibble(paths2) 
R> paths2_df %>% head()
\end{CodeInput}

\begin{CodeOutput}
# A tibble: 6 x 8
  .node .parent .depth outcomes next_dose fit       parent_fit dose_index
  <dbl>   <dbl>  <dbl> <chr>        <dbl> <list>    <list>     <list>    
1     1      NA      0 ""               2 <crm_fit> <NULL>     <dbl [5]> 
2     2       1      1 NNN              3 <crm_fit> <crm_fit>  <dbl [5]> 
3     3       2      2 NNN              4 <crm_fit> <crm_fit>  <dbl [5]> 
4     4       1      1 NNT              2 <crm_fit> <crm_fit>  <dbl [5]> 
5     5       4      2 NNN              3 <crm_fit> <crm_fit>  <dbl [5]> 
6     6       1      1 NTT              1 <crm_fit> <crm_fit>  <dbl [5]> 
\end{CodeOutput}
\end{CodeChunk}

Each node in the graph has a single row in this table. The variables
that define the tree structure are prefixed with a period. Whilst this
tall format is useful for some purposes, like creating graphs like
Figure \ref{fig:crm_dtps_graph}, dose-transition pathways are perhaps
most intuitively tabulated in a wide format. That task is accomplished
by the \texttt{spread\_paths} function, that will join \texttt{.node} to
\texttt{.parent} at each \texttt{.depth} to create a wide view of DTPs.
For example,

\begin{CodeChunk}

\begin{CodeInput}
R> spread_paths(paths2_df %>% select(-fit, -parent_fit, -dose_index))
\end{CodeInput}

\begin{CodeOutput}
# A tibble: 16 x 6
   outcomes0 next_dose0 outcomes1 next_dose1 outcomes2 next_dose2
   <chr>          <dbl> <chr>          <dbl> <chr>          <dbl>
 1 ""                 2 NNN                3 NNN                4
 2 ""                 2 NNN                3 NNT                3
 3 ""                 2 NNN                3 NTT                2
 4 ""                 2 NNN                3 TTT                2
 5 ""                 2 NNT                2 NNN                3
 6 ""                 2 NNT                2 NNT                2
 7 ""                 2 NNT                2 NTT                1
 8 ""                 2 NNT                2 TTT                1
 9 ""                 2 NTT                1 NNN                2
10 ""                 2 NTT                1 NNT                1
11 ""                 2 NTT                1 NTT                1
12 ""                 2 NTT                1 TTT                1
13 ""                 2 TTT                1 NNN                1
14 ""                 2 TTT                1 NNT                1
15 ""                 2 TTT                1 NTT                1
16 ""                 2 TTT                1 TTT               NA
\end{CodeOutput}
\end{CodeChunk}

In this view, paths can be read horizontally. This confirms the dose
choices shown in Figure \ref{fig:crm_dtps_graph}. Had we retained
\texttt{fit} in the frame that was passed to \texttt{spread\_paths},
there would also be columns \texttt{fit0}, \texttt{fit1} and
\texttt{fit2} in the table above. These functions create a rich and
flexible method for working with paths in CRM trials. We demonstrate a
similar approach in the next section with EffTox.

Fitting TITE-CRM models is also supported in \pkg{trialr}, again through
the \texttt{stan\_crm} function. The outcome syntax used above, however,
ceases to be succinct and useful when each patient is associated an
individual weight. Thus, when fitting the TITE-CRM, we provide vectors
for the doses given, the binary toxicity outcomes observed, and the
weight for each observation. For instance, in the example below we
calculate the next dose in a trial that has partially evaluated four
patients at dose 3, as described by \citet[p.124]{Cheung2011}. No
patient has yet experienced DLT but they are between 28 and 73 days
through a 126-day evaluation window.

\begin{CodeChunk}

\begin{CodeInput}
R> fit <-stan_crm(skeleton = c(0.05, 0.12, 0.25, 0.40, 0.55), target = 0.25,
R+                doses_given = c(3, 3, 3, 3),
R+                tox = c(0, 0, 0, 0),
R+                weights = c(73, 66, 35, 28) / 126,
R+                model = 'empiric', beta_sd = sqrt(1.34), 
R+                seed = 123, refresh = 0)
R> fit$recommended_dose
\end{CodeInput}

\begin{CodeOutput}
[1] 4
\end{CodeOutput}
\end{CodeChunk}

We receive confirmation that the model advocates starting the next
patient on dose 4.

\hypertarget{efftox-1}{%
\subsection{EffTox}\label{efftox-1}}

The EffTox model addresses the need for a dose-finding design that
incorporates both efficacy and toxicity outcomes. This is useful in
modern oncology treatments like small molecules and immunotherapies
where, compared to chemotherapy, there is a weaker rationale for the
assumption that higher doses are associated with higher probabilities of
success.

The probability model for EffTox described above uses six parameters.
The extra complexity naturally means it is more challenging to select
priors. \citet{Thall2014} describe an optimisation algorithm that
selects hyperparameters for weakly informative normal priors on
\(\alpha, \beta, \gamma\) and \(\zeta\) so that the expected prior
probabilities of efficacy and toxicity at each dose match
investigator-elicited values. The priors for \(\eta\) and \(\psi\) are
fixed at \(N(0, 0.2)\) and \(N(0, 1)\) respectively.

For example, expecting efficacy probabilities (0.2, 0.4, 0.6, 0.8, 0.9)
and toxicity probabilities (0.02, 0.04, 0.06, 0.08, 0.10) in a five-dose
example, \citet{Thall2014} choose the parameter priors in Table
\ref{tab:efftox_priors}. The hyperparameter values can be calculated
using the MD Anderson EffTox software. This algorithm is not currently
implemented in \pkg{trialr}.

\begin{table}
  \centering
  \begin{tabular}{| l | l | l |}
  \hline
   Parameter &  Mean &  Standard deviation \\ 
   \hline
   $\alpha$ & -7.9593 & 3.5487 \\
   $\beta$ & 1.5482 & 3.5018 \\
   $\gamma$ & 0.7367 & 2.5423 \\ 
   $\zeta$ & 3.4181 & 2.4406 \\
   $\eta$ & 0 & 0.2 \\
   $\psi$ & 0 & 1 \\
   \hline
  \end{tabular}
  \caption{Normal prior hyperparameters used by \citet{Thall2014}.}
  \label{tab:efftox_priors}
\end{table}

\begin{CodeChunk}
\begin{figure}

{\centering \includegraphics{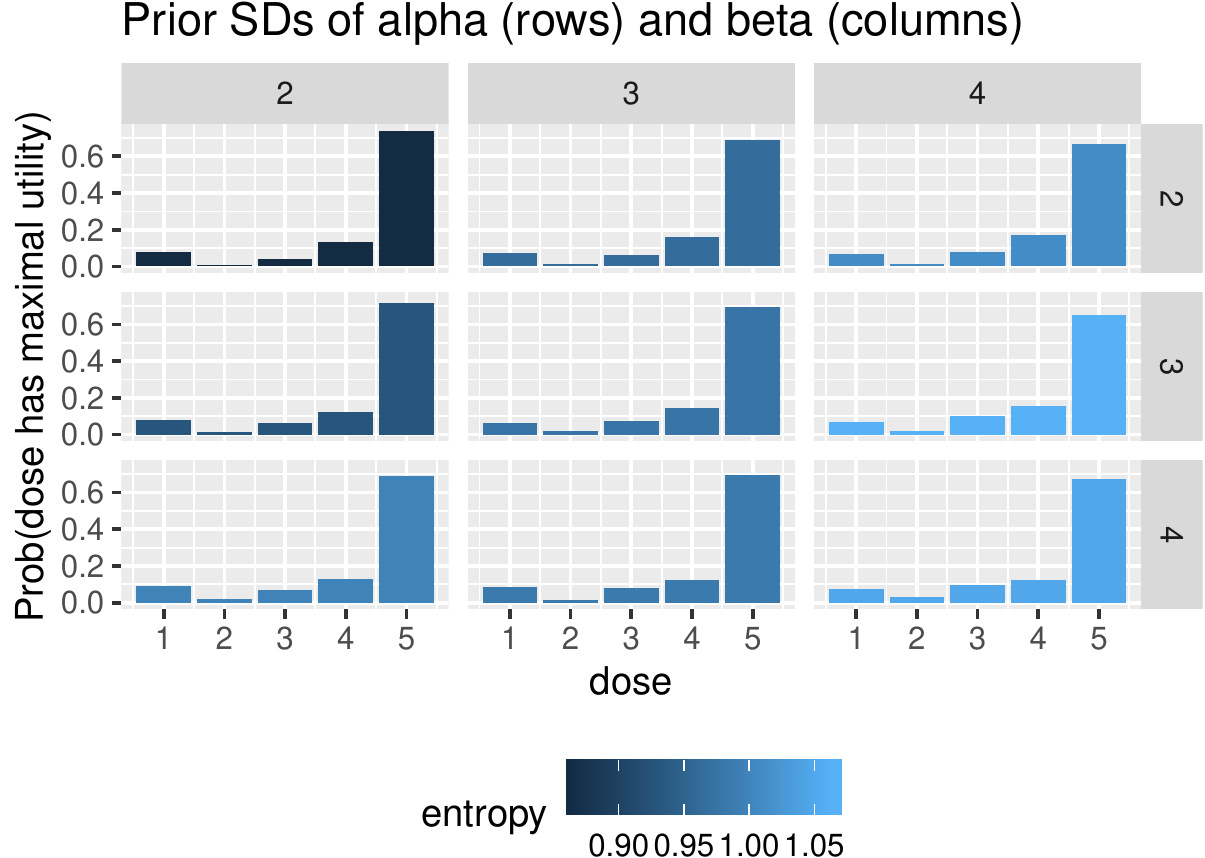} 

}

\caption[The prior standard deviations on alpha and beta in EffTox have relatively weak effects on the prior beliefs on the location of optimal dose]{The prior standard deviations on alpha and beta in EffTox have relatively weak effects on the prior beliefs on the location of optimal dose.}\label{fig:efftox_prior_plot_1}
\end{figure}
\end{CodeChunk}

\begin{CodeChunk}
\begin{figure}

{\centering \includegraphics{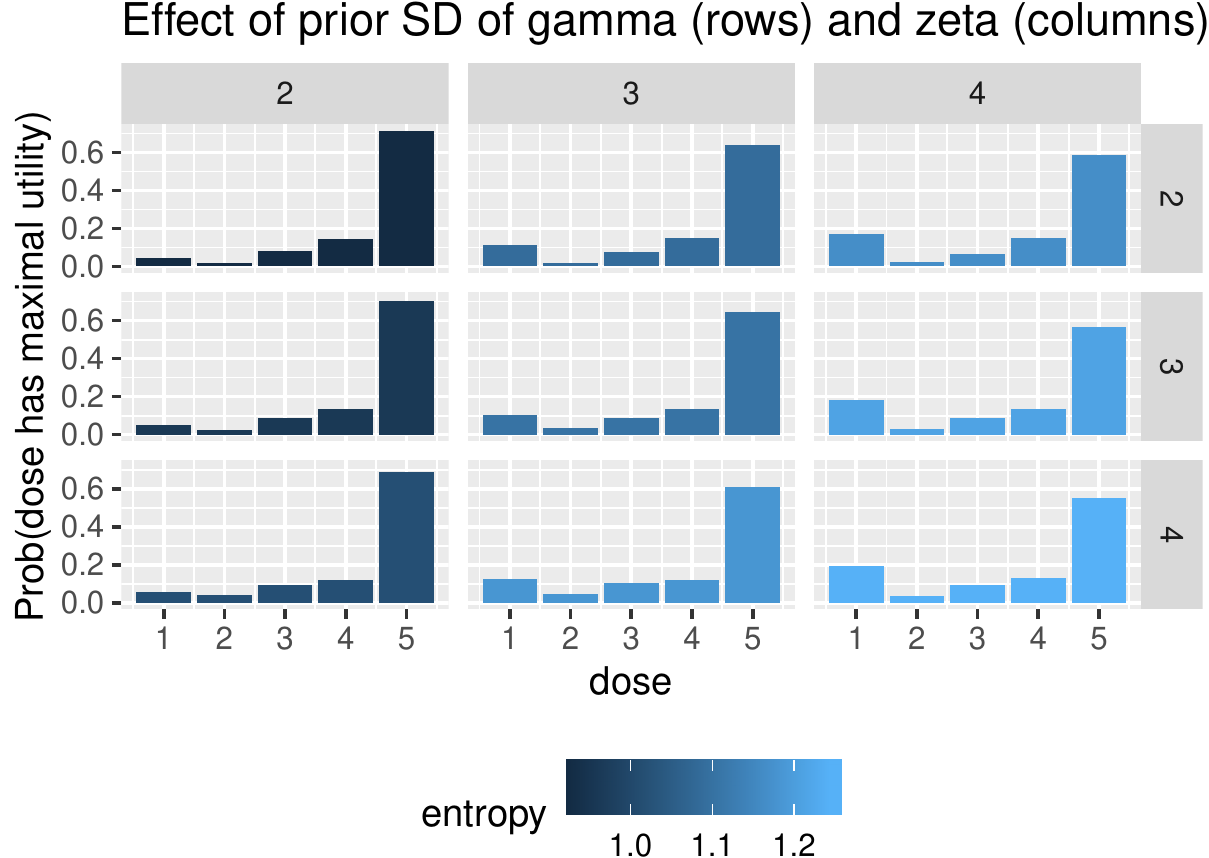} 

}

\caption[The prior standard deviations on gamma and zeta in EffTox have relatively weak effects on the prior beliefs on the location of optimal dose]{The prior standard deviations on gamma and zeta in EffTox have relatively weak effects on the prior beliefs on the location of optimal dose.}\label{fig:efftox_prior_plot_2}
\end{figure}
\end{CodeChunk}

Figures \ref{fig:efftox_prior_plot_1} and \ref{fig:efftox_prior_plot_2}
show that the prior views on the location of the optimal dose do not
change materially for modest changes in the values of the standard
deviation hyperparameters. In each, the value of the non-varying
parameters takes the value listed in Table \ref{tab:efftox_priors}. We
see in each panel of each plot that dose-level 5 is overwhelmingly
expected to have the highest utility score. As the standard deviation
hyperparameters increase, more probability is placed on dose 1 being
optimal. The probabilities of doses 2 to 4 barely change.

By way of illustration, let us continue with Thall \textit{et al.}'s
parameterisation. The authors seek the most attractive dose of a drug
from (1, 2, 4, 6.6, 10) mcL/kg that satisfies both

\begin{equation}
Pr(\pi_E(x_i) > 0.5 | \mathcal{D}) > 0.1
\end{equation}

and

\begin{equation}
Pr(\pi_T(x_i) < 0.3 | \mathcal{D}) > 0.1
\end{equation}

where the neutral utility contour intersects the hinge points
\((\pi_{1,E}^*, 0) = (0.5, 0)\), \((1, \pi_{2,T}^*) = (1, 0.65)\) and
\((\pi_{3,E}^*, \pi_{3,T}^*) = (0.7, 0.25)\). Let us fit this model to
the outcomes \texttt{1NNN\ 2ENN}:

\begin{CodeChunk}

\begin{CodeInput}
R> outcomes <- '1NNN 2ENN'
R> 
R> fit <- stan_efftox(outcome_str = outcomes,
R+                    real_doses = c(1.0, 2.0, 4.0, 6.6, 10.0),
R+                    efficacy_hurdle = 0.5, toxicity_hurdle = 0.3,
R+                    p_e = 0.1, p_t = 0.1, eff0 = 0.5, tox1 = 0.65,
R+                    eff_star = 0.7, tox_star = 0.25,
R+                    alpha_mean = -7.9593, alpha_sd = 3.5487,
R+                    beta_mean = 1.5482, beta_sd = 3.5018,
R+                    gamma_mean = 0.7367, gamma_sd = 2.5423,
R+                    zeta_mean = 3.4181, zeta_sd = 2.4406,
R+                    eta_mean = 0, eta_sd = 0.2,
R+                    psi_mean = 0, psi_sd = 1, 
R+                    seed = 123, refresh = 0)
R> fit
\end{CodeInput}

\begin{CodeOutput}
  Patient Dose Toxicity Efficacy
1       1    1        0        0
2       2    1        0        0
3       3    1        0        0
4       4    2        0        1
5       5    2        0        0
6       6    2        0        0

  Dose N ProbEff ProbTox ProbAccEff ProbAccTox Utility Acceptable ProbOBD
1    1 3  0.0506 0.00685      0.003      0.997  -0.911      FALSE  0.0158
2    2 3  0.2700 0.00356      0.130      1.000  -0.466       TRUE  0.0075
3    3 0  0.7287 0.01399      0.798      0.988   0.434       TRUE  0.0862
4    4 0  0.8667 0.05261      0.911      0.940   0.648      FALSE  0.1893
5    5 0  0.9121 0.11791      0.938      0.861   0.637      FALSE  0.7013

The model recommends selecting dose-level 3.
The dose most likely to be the OBD is 5.
Model entropy: 0.88
\end{CodeOutput}
\end{CodeChunk}

The parameters \texttt{eff0}, \texttt{tox1}, \texttt{eff\_star} and
\texttt{tox\_star} reflect the values \(\pi_{1,E}^*\), \(\pi_{2,T}^*\),
\((\pi_{3,E}^*\) and \(\pi_{3,T}^*\).

We see that dose-level 3 is recommended for the next cohort. Even though
doses 4 and 5 have higher estimated utility scores, they are not
recommended because dose 3 has not yet been given. \texttt{ProbAccEff}
and \texttt{ProbAccTox} estimate the probability that the estimated
rates of efficacy and toxicity are the desirable sides of 0.5 and 0.3,
respectively. Dose 1 is not acceptable because \texttt{ProbAccEff} is
less than 0.1. The utility values implied by the posterior mean
probabilities of efficacy and toxicity are shown in Figure
\ref{fig:efftox_contours}, created using the command:

\begin{CodeChunk}

\begin{CodeInput}
R> efftox_contour_plot(fit)
\end{CodeInput}
\begin{figure}

{\centering \includegraphics{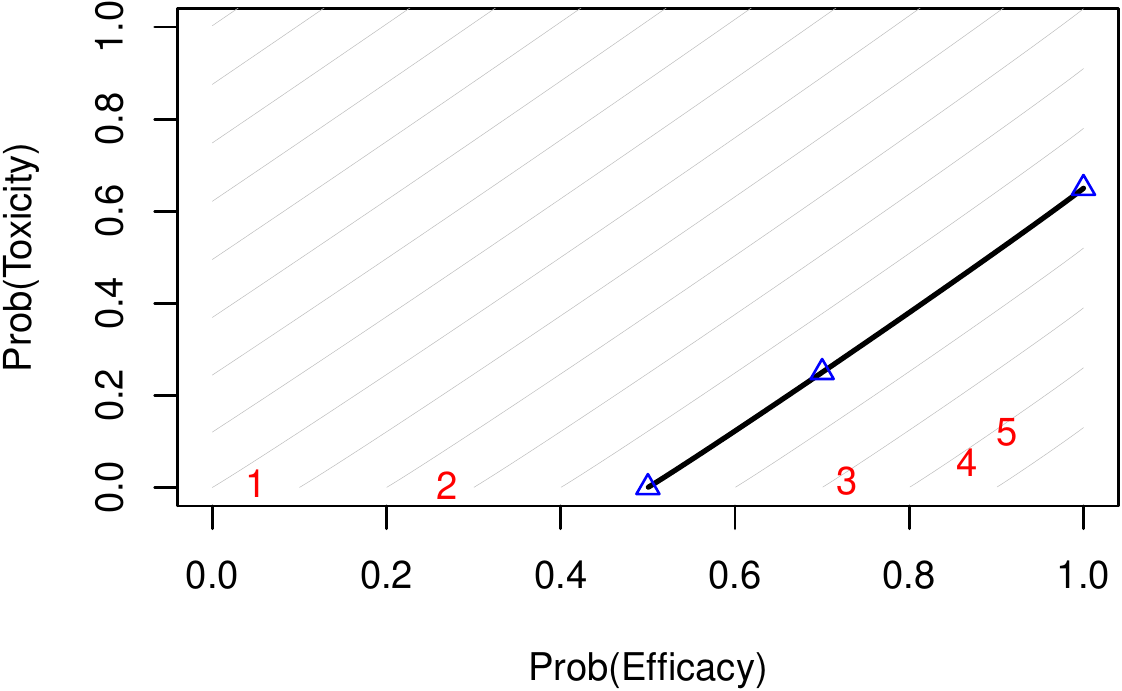} 

}

\caption[Utility contour plot for the described EffTox example after observing 1NNN 2ENN]{Utility contour plot for the described EffTox example after observing 1NNN 2ENN.}\label{fig:efftox_contours}
\end{figure}
\end{CodeChunk}

Doses that are closer to the lower-right corner of Figure
\ref{fig:efftox_contours}, reflecting guarenteed efficacy and no
toxicity, have higher utility scores.

Using the posterior samples of the utility scores at each dose, it is
possible to estimate the probability that each dose is superior to the
others. The following superiority matrix shows the implied probability
that the dose represented by the column has greater utility than the
dose represented by the row:

\begin{CodeChunk}

\begin{CodeInput}
R> efftox_superiority(fit)
\end{CodeInput}

\begin{CodeOutput}
        1       2       3       4       5
1      NA 0.98125 0.97800 0.97350 0.96800
2 0.01875      NA 0.97575 0.96075 0.93150
3 0.02200 0.02425      NA 0.89200 0.81300
4 0.02650 0.03925 0.10800      NA 0.70825
5 0.03200 0.06850 0.18700 0.29175      NA
\end{CodeOutput}
\end{CodeChunk}

Thus, we can be quite sure that dose 4 has higher utility than doses 1
to 3, but we are rather more unsure about whether it has higher utility
than dose 5. This is confirmed by Figure \ref{fig:efftox_contours}.

\pkg{trialr} also supports the calculation of dose-transition pathways
for EffTox. The number of events that patients may experience is now
four (E, T, B or N), compared to the two events (T or N) that were
possible in the CRM examples. As such, the number of distinct pathways
that can be taken increases very quickly. For instance, the number of
outcome combinations for a cohort of three is 20, so that two cohorts of
three generate 400 pathways. This is too many to represent on a graph.
To illustrate, we consider the next cohort of three patients in our
trial above:

\begin{CodeChunk}

\begin{CodeInput}
R> paths <- efftox_dtps(cohort_sizes = c(3), previous_outcomes = outcomes, 
R+                      real_doses = c(1.0, 2.0, 4.0, 6.6, 10.0),
R+                      efficacy_hurdle = 0.5, toxicity_hurdle = 0.3,
R+                      p_e = 0.1, p_t = 0.1, eff0 = 0.5, tox1 = 0.65,
R+                      eff_star = 0.7, tox_star = 0.25,
R+                      alpha_mean = -7.9593, alpha_sd = 3.5487,
R+                      beta_mean = 1.5482, beta_sd = 3.5018,
R+                      gamma_mean = 0.7367, gamma_sd = 2.5423,
R+                      zeta_mean = 3.4181, zeta_sd = 2.4406,
R+                      eta_mean = 0, eta_sd = 0.2,
R+                      psi_mean = 0, psi_sd = 1, 
R+                      next_dose = 3, seed = 123, refresh = 0)
\end{CodeInput}
\end{CodeChunk}

Again, the returned object has type \texttt{dose\_finding\_paths} so, as
with the CRM examples, we can invoke \texttt{as\_tibble} and
\texttt{spread\_paths} on \texttt{paths}. The associated \texttt{fit}
and \texttt{parent\_fit} objects are each of type \texttt{efftox\_fit}.
These can be used with \texttt{map}-like functions from \pkg{purrr}
\citep{purrr} for flexible analysis of DTP objects:

\begin{CodeChunk}

\begin{CodeInput}
R> library(tidyr)
R> library(purrr)
R> 
R> as_tibble(paths) %>% 
R+   filter(.depth > 0) %>% 
R+   mutate(prob_obd = map(fit, 'prob_obd'), 
R+          parent_prob_obd = map(parent_fit, 'prob_obd')) %>% 
R+   select(outcomes, dose_index, prob_obd, parent_prob_obd) %>% 
R+   unnest %>% 
R+   mutate(prob_obd_delta = prob_obd - parent_prob_obd) %>% 
R+   filter(dose_index == 5)
\end{CodeInput}

\begin{CodeOutput}
# A tibble: 20 x 5
   outcomes dose_index prob_obd parent_prob_obd prob_obd_delta
   <chr>         <int>    <dbl>           <dbl>          <dbl>
 1 BBB               5    0.136           0.701        -0.565 
 2 BBE               5    0.061           0.701        -0.640 
 3 BBN               5    0.176           0.701        -0.525 
 4 BBT               5    0.316           0.701        -0.385 
 5 BEE               5    0.142           0.701        -0.559 
 6 BEN               5    0.21            0.701        -0.491 
 7 BET               5    0.164           0.701        -0.538 
 8 BNN               5    0.323           0.701        -0.378 
 9 BNT               5    0.274           0.701        -0.427 
10 BTT               5    0.397           0.701        -0.305 
11 EEE               5    0.677           0.701        -0.0245
12 EEN               5    0.756           0.701         0.0550
13 EET               5    0.223           0.701        -0.478 
14 ENN               5    0.788           0.701         0.0872
15 ENT               5    0.312           0.701        -0.389 
16 ETT               5    0.280           0.701        -0.421 
17 NNN               5    0.728           0.701         0.0272
18 NNT               5    0.329           0.701        -0.372 
19 NTT               5    0.242           0.701        -0.460 
20 TTT               5    0.287           0.701        -0.415 
\end{CodeOutput}
\end{CodeChunk}

In the example above for instance, we illustrate how the views on
dose-level 5, in particular its chances of being the optimal dose, can
change materially based on the outcome of the next cohort of three
patients. This demonstrates the lack of information currently in the
model in this situation, particularly the sensitivity of the estimated
identity of the OBD to incidence of any toxicity.

\begin{figure}
  \centering
  \includegraphics[width=0.45\columnwidth]{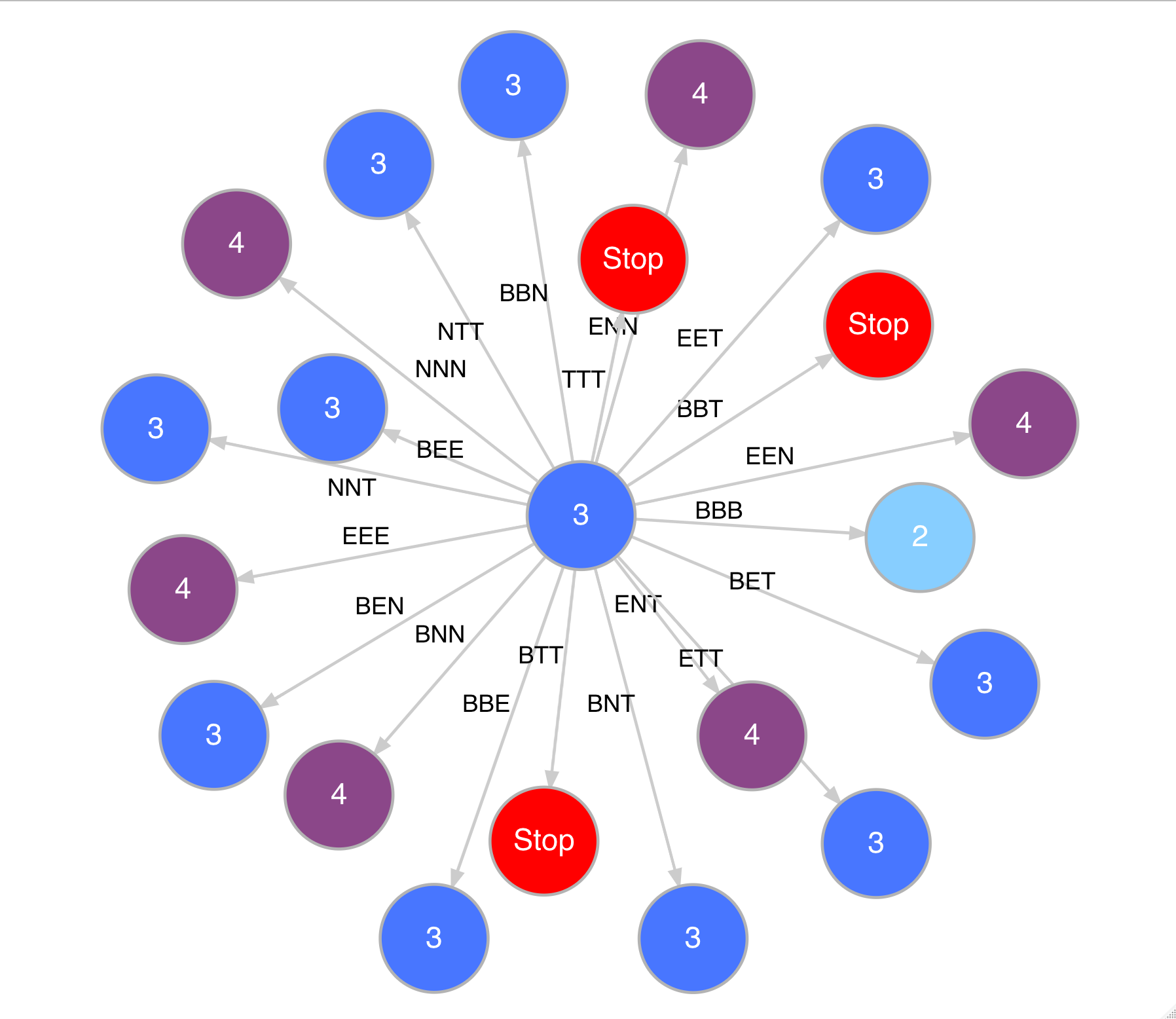}
  \includegraphics[width=0.45\columnwidth]{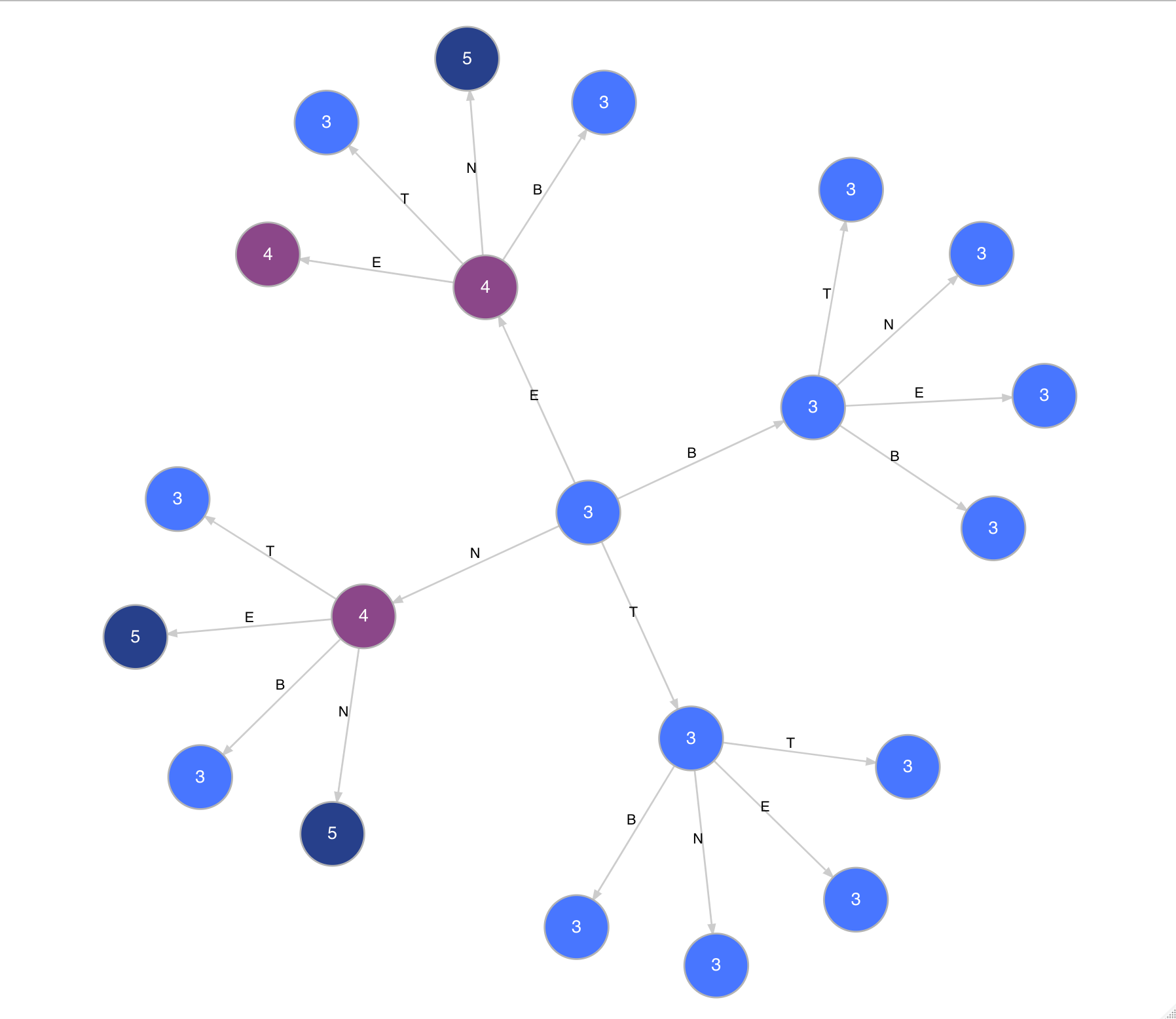}
  \caption{Dose pathways for the next cohort of three patients (left) and next two cohorts of one patient (right) in an EffTox trial that has already seen `1NNN 2ENN`.}
  \label{fig:efftox_dtps_graph}
\end{figure}

The left panel of Figure \ref{fig:efftox_dtps_graph} shows the paths for
the next cohort of three. After only nine patients, there are three
paths that advocate stopping. These are all paths that see considerable
toxicity in cohort 3. The right panel of Figure
\ref{fig:efftox_dtps_graph} shows the paths generated by the next two
cohorts of one patient. Code to create these graphs appears in the
Appendix.

The function \texttt{efftox\_dtps} also accepts a custom dose selection
function via the \texttt{user\_dose\_func} parameter. The \texttt{fit}
object recorded at each node in \texttt{paths} has type
\texttt{efftox\_fit}, a subclass of \texttt{dose\_finding\_fit}. The
\texttt{user\_dose\_func} delegate should take an \texttt{efftox\_fit}
as the sole parameter and return an integer dose-level for the next
patient, or \texttt{NA} to advocate stopping the trial, mirroring the
behaviour of the CRM example. We do not illustrate that here.

\hypertarget{augmented-binary-method-1}{%
\subsection{Augmented Binary method}\label{augmented-binary-method-1}}

To demonstrate the two-period Augmented Binary method for single arm
trials, we will fit it to a dataset drawn from one of the simulation
scenarios used by \citet{Wason2013}. First, let us consider priors on
the parameters.

We will consider two sets of priors to demonstrate the effect they have
on the posterior. The prior distributions in the first set are
relatively informative:

\begin{CodeChunk}

\begin{CodeInput}
R> informative_priors <- list(alpha_mean = 0, alpha_sd = 0.1,
R+                            beta_mean = 0, beta_sd = 0.1,
R+                            gamma_mean = 0, gamma_sd = 0.1,
R+                            sigma_mean = 0, sigma_sd = 0.5,
R+                            omega_lkj_eta = 1,
R+                            alpha_d1_mean = 0, alpha_d1_sd = 0.5,
R+                            gamma_d1_mean = 0, gamma_d1_sd = 0.25,
R+                            alpha_d2_mean = 0, alpha_d2_sd = 0.5,
R+                            gamma_d2_mean = 0, gamma_d2_sd = 0.25)
\end{CodeInput}
\end{CodeChunk}

For the covariance matrix \(\Sigma\), half-normal priors with parameters
\texttt{sigma\_mean} and \texttt{sigma\_sd} are used for the diagonal
elements, and an LKJ prior is specified on the associated correlation
matrix, \(\Omega\). The parameter \(\Omega_\eta = 1\) generates uniform
beliefs on the correlation between \(y_1\) and \(y_2\). Values \(> 1\)
would reflect an expectation that correlations near 0 are more likely,
and vice-versa.

In constrast, the priors in the second set are relatively diffuse:

\begin{CodeChunk}

\begin{CodeInput}
R> diffuse_priors <- list(alpha_mean = 0, alpha_sd = 1,
R+                        beta_mean = 0, beta_sd = 1,
R+                        gamma_mean = 0, gamma_sd = 1,
R+                        sigma_mean = 0, sigma_sd = 1,
R+                        omega_lkj_eta = 1,
R+                        alpha_d1_mean = 0, alpha_d1_sd = 1,
R+                        gamma_d1_mean = 0, gamma_d1_sd = 1,
R+                        alpha_d2_mean = 0, alpha_d2_sd = 1,
R+                        gamma_d2_mean = 0, gamma_d2_sd = 1)
\end{CodeInput}
\end{CodeChunk}

Unit normal prior distributions may not immediately seem diffuse until
we consider their effect on the generated outcome distributions. Prior
predictive outcomes generated by each set of priors are shown in Figures
\ref{fig:augbin_prior_pred_tumour} and \ref{fig:augbin_prior_pred_nsf}.
Code to create these figures are included in the Appendix.

\begin{CodeChunk}
\begin{figure}

{\centering \includegraphics{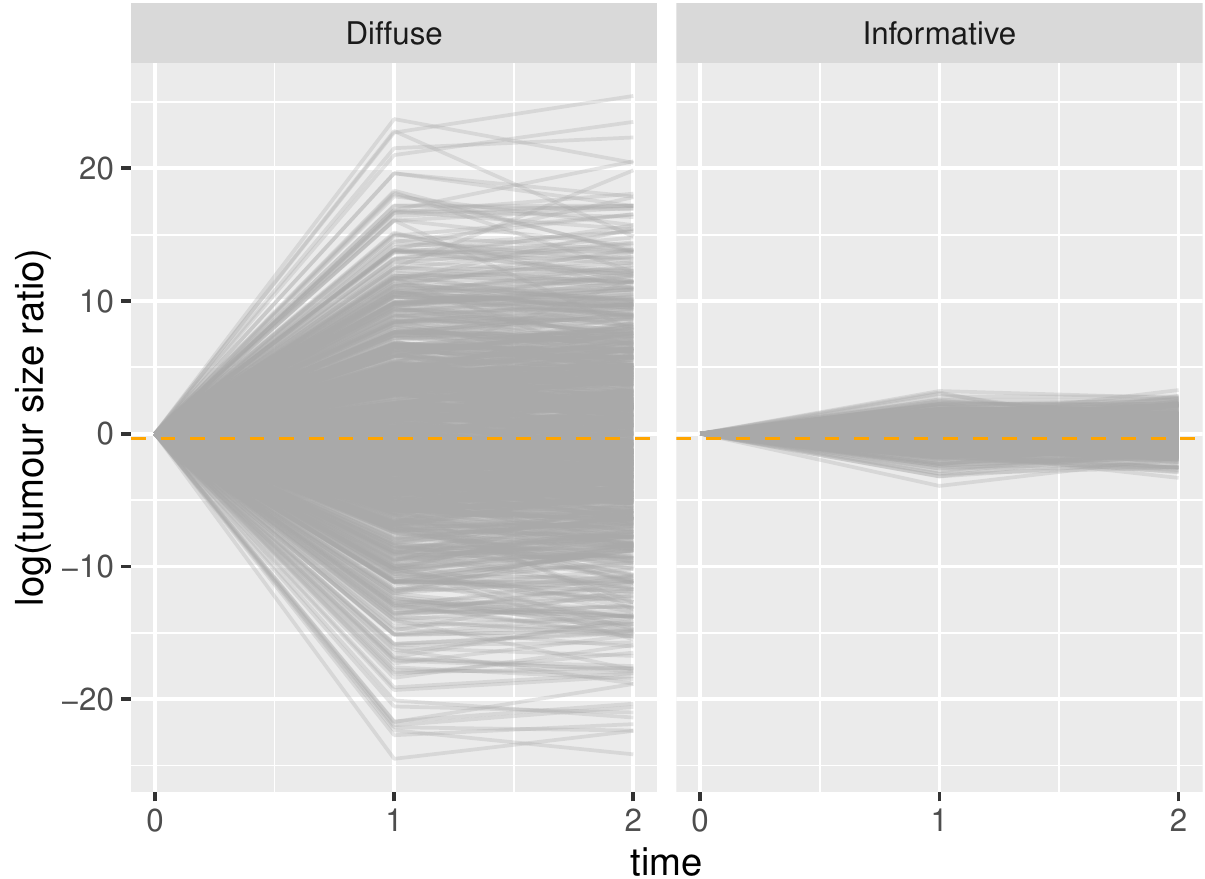} 

}

\caption[Prior predictive tumour size ratios generated by two sets of prior parameters in the two-period AugBin model]{Prior predictive tumour size ratios generated by two sets of prior parameters in the two-period AugBin model.}\label{fig:augbin_prior_pred_tumour}
\end{figure}
\end{CodeChunk}

We see in Figure \ref{fig:augbin_prior_pred_tumour} that the diffuse
priors generate a far wider set of outcomes. In fact, log-tumour-size
ratios near 20 imply tumour growth that is many orders of magnitude
greater than what is feasible. Growth to this extent would imply that
the patient could increase their volume thousands of times over, a clear
nonsense. The outcomes generated by the informative priors are much more
realistic. The baseline tumour sizes in cm are sampled from a U(5, 10)
distribution so that the expected tumour size in each set of priors is
7.5cm. Under the informative priors, the expected tumour size at each
post-baseline assessment is 12.5cm. Under each set, the chances of
tumour growth at each timepoint is roughly 50\%.

\begin{CodeChunk}
\begin{figure}

{\centering \includegraphics{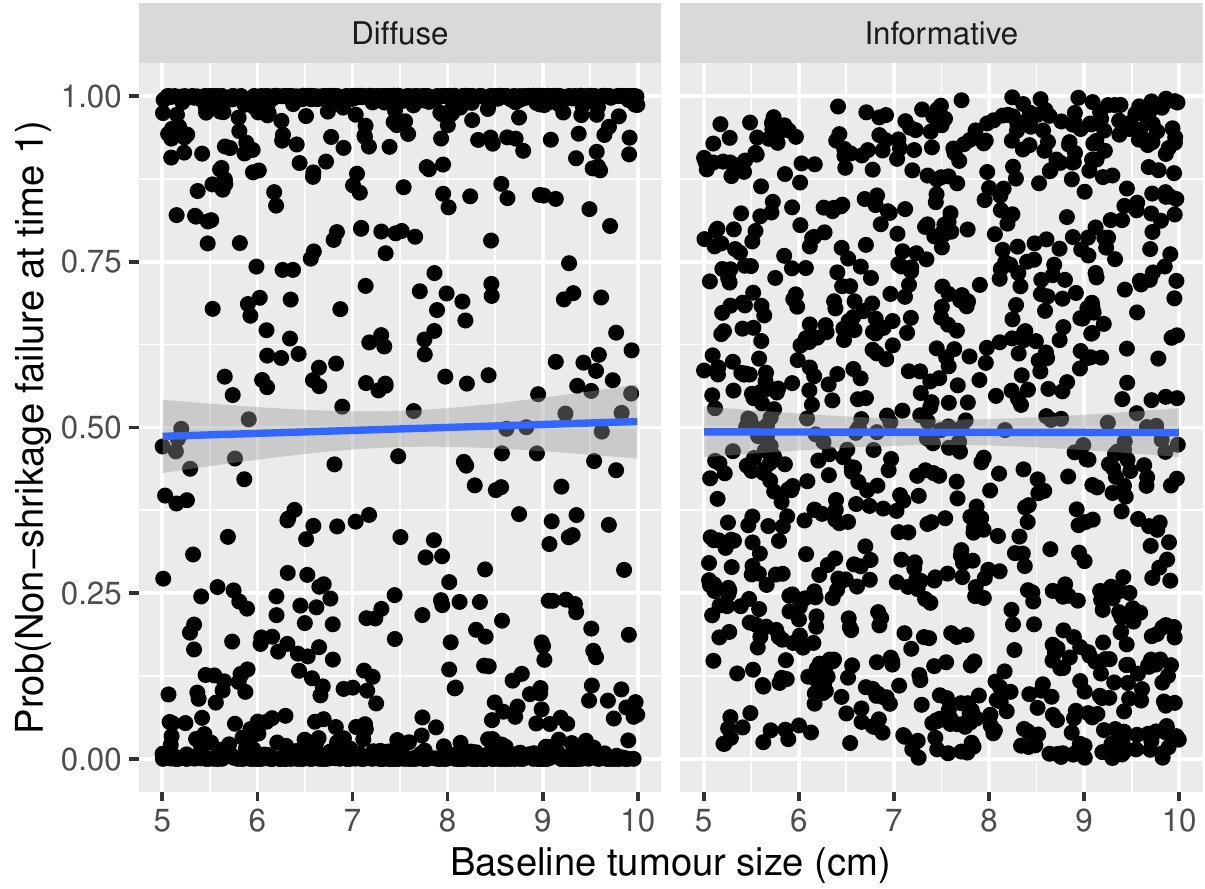} 

}

\caption[Prior predictive probabilities of non-shrinkage failure generated by two sets of prior parameters in the two-period AugBin model]{Prior predictive probabilities of non-shrinkage failure generated by two sets of prior parameters in the two-period AugBin model.}\label{fig:augbin_prior_pred_nsf}
\end{figure}
\end{CodeChunk}

Figure \ref{fig:augbin_prior_pred_nsf} shows the prior predictive
probabilities of non-shrinkage failure generated by the two prior sets
at the interim timepoint. Once again, we see that extreme outcomes are
generated far too readily by the diffuse priors. In contrast, the
informative priors generate probabilities that are roughly uniform and
not associated with baseline tumour size.

To demonstrate the AugBin model, we will fit it to a dataset randomly
sampled using the parameterisation described by \citet{Wason2013} in the
first line of their Table 1. We have parameters:

\begin{CodeChunk}

\begin{CodeInput}
R> N <- 50
R> sigma <- 1
R> delta1 <- -0.356
R> mu <- c(0.5 * delta1, delta1)
R> Sigma = matrix(c(0.5 * sigma^2, 0.5 * sigma^2, 
R+                  0.5 * sigma^2, sigma^2), ncol = 2)
R> alphaD <- -1.5
R> gammaD <- 0
\end{CodeInput}
\end{CodeChunk}

That is, there are 50 patients, the expected tumour shrinkage at time 2
is \(\exp(-0.356) - 1 = 30\%\), and the chances of non-shrinkage failure
at each stage are \(\text{logit}^{-1}(-1.5) = 18\%\). These parameters
generate data:

\begin{CodeChunk}

\begin{CodeInput}
R> set.seed(123456)
R> y <- MASS::mvrnorm(n = N, mu, Sigma)
R> z0 <- runif(N, min = 5, max = 10)
R> z1 <- exp(y[, 1]) * z0
R> z2 <- exp(y[, 2]) * z0
R> d1 <- rbinom(N, size = 1, prob = gtools::inv.logit(alphaD + gammaD * z0))
R> d2 <- rbinom(N, size = 1, prob = gtools::inv.logit(alphaD + gammaD * z1))
R> tumour_size = data.frame(z0, z1, z2) # cm
R> non_shrinkage_failure <- data.frame(d1, d2)
\end{CodeInput}
\end{CodeChunk}

The values in \texttt{tumour\_size} reflect \((z_1, z_2, z_3)\) measured
in cm.

To fit this data to the two-period, single-arm model, referred to as
\texttt{2t-1a} in \pkg{trialr}, we run the command:

\begin{CodeChunk}

\begin{CodeInput}
R> fit_diffuse <- stan_augbin(tumour_size, non_shrinkage_failure, 
R+                            model = '2t-1a', prior_params = diffuse_priors, 
R+                            seed = 123, refresh = 0)
R> fit_inf <- stan_augbin(tumour_size, non_shrinkage_failure, 
R+                        model = '2t-1a', prior_params = informative_priors, 
R+                        seed = 123, refresh = 0)
\end{CodeInput}
\end{CodeChunk}

The object returned by \texttt{stan\_fit} has type \texttt{augbin\_fit}.

The main benefit that Wason \& Seaman identify for their method is the
increase in efficiency that comes from analysing the continuous tumour
size variable rather than dichotomising the information into a binary
outcome. They calculate the width of the 95\% confidence interval (CI)
for the probability of success under their method. They then do the same
using the dichotomised binary success variables, \(S_i\), using 30\%
shrinkage at time 2 and no non-shrinkage failure as the criteria for
success. They infer the increase in efficiency by comparing the widths
of the CI under each method from a large number of simulated trials. We
can recreate this type of analysis in \pkg{trialr}. Focusing on the
model fit using the diffuse priors:

\begin{CodeChunk}

\begin{CodeInput}
R> pred_diffuse <- predict(fit_diffuse, y2_upper = log(0.7))
R> pred_diffuse %>% head()
\end{CodeInput}

\begin{CodeOutput}
# A tibble: 6 x 8
     id    z0    z1 prob_success_samp prob_success lower upper ci_width
  <int> <dbl> <dbl> <list>                   <dbl> <dbl> <dbl>    <dbl>
1     1  5.99 16.0  <dbl [4,000]>            0.277 0.176 0.385    0.208
2     2  5.66  4.81 <dbl [4,000]>            0.223 0.141 0.318    0.177
3     3  9.32  5.59 <dbl [4,000]>            0.269 0.165 0.385    0.220
4     4  6.16  3.16 <dbl [4,000]>            0.218 0.136 0.316    0.180
5     5  5.57 10.2  <dbl [4,000]>            0.250 0.161 0.352    0.191
6     6  7.08 11.1  <dbl [4,000]>            0.275 0.186 0.374    0.188
\end{CodeOutput}
\end{CodeChunk}

We see that the estimated probabilities of success are approximately
25\%, and vary slightly depending on the baseline and interim tumour
size measurements. The \texttt{prob\_success\_samp} column contains the
posterior samples of the probability of success for each patient. By
default, \texttt{lower} and \texttt{upper} contain posterior 2.5\% and
97.5\% quantiles of \texttt{prob\_success} but this can be changed by
specifying the desired quantile probabilities using the \texttt{probs}
argument.

We can also make inferences by analysing the dichotomised binary
outcome:

\begin{CodeChunk}

\begin{CodeInput}
R> pred_binary <- binary_prob_success(fit_diffuse, y2_upper = log(0.7), 
R+                                    methods = 'exact')
R> pred_binary
\end{CodeInput}

\begin{CodeOutput}
  method  x  n mean     lower     upper  ci_width
1  exact 14 50 0.28 0.1623106 0.4249054 0.2625948
\end{CodeOutput}
\end{CodeChunk}

This function uses the \pkg{binom} \citep{binom} package.

We see that there were 14 responses in 50 patients, yielding a response
rate of 28\%. By default, bounds for the 95\% CI are shown but this can
be tailored. The average decrease in the widths of the 95\% intervals
is:

\begin{CodeChunk}

\begin{CodeInput}
R> mean(pred_diffuse$ci_width) / pred_binary$ci_width - 1
\end{CodeInput}

\begin{CodeOutput}
[1] -0.2262706
\end{CodeOutput}
\end{CodeChunk}

The reduction of approximately 23\% in this single simulated iteration
is higher than the 16.5\% average reduction reported by Wason \& Seaman
comparing AugBin to the regular binary method in this scenario.

\begin{CodeChunk}
\begin{figure}

{\centering \includegraphics{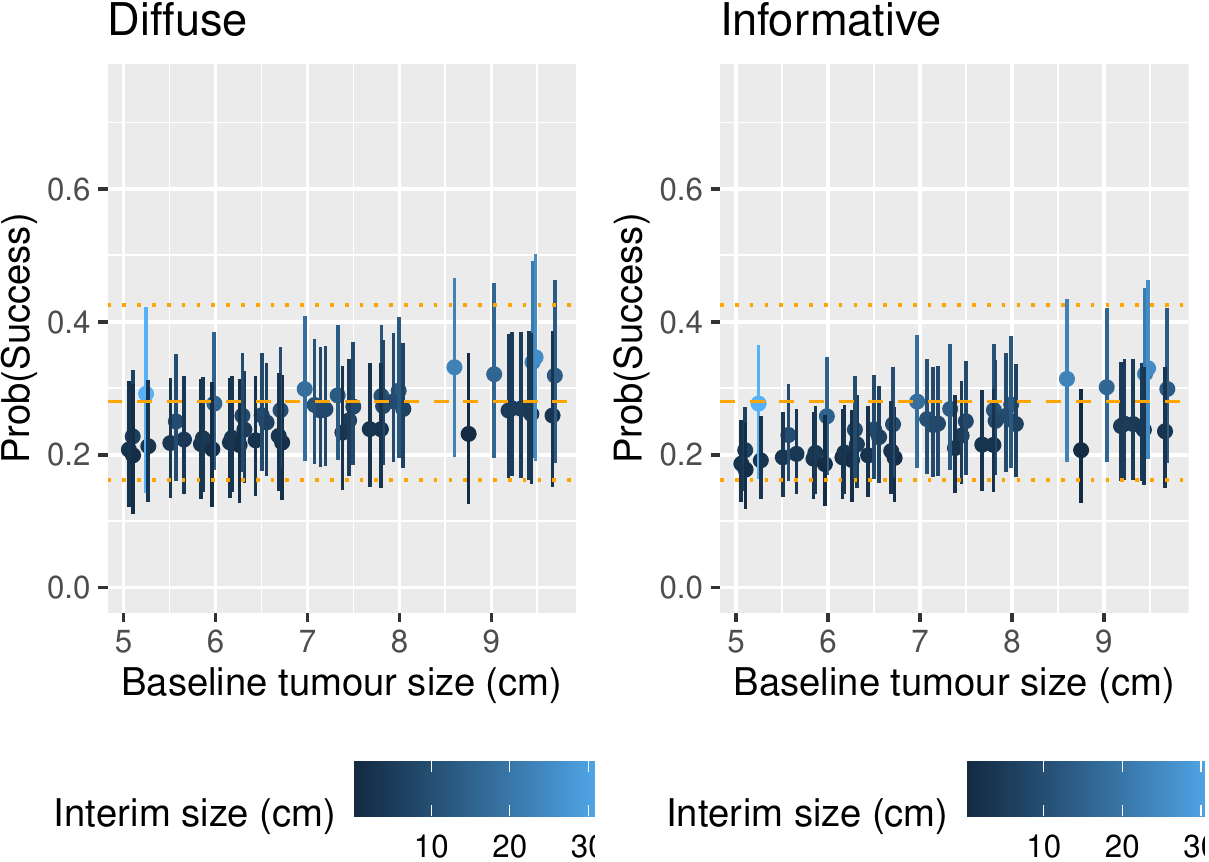} 

}

\caption[Posterior predictions from the AugBin model to 50 simulated patients under two different sets of priors]{Posterior predictions from the AugBin model to 50 simulated patients under two different sets of priors. The orange lines show the mean and 0.95 CI derived from the binary success variables.}\label{fig:post_pred_augbin}
\end{figure}
\end{CodeChunk}

These inferences for both AugBin fits are shown in Figure
\ref{fig:post_pred_augbin}. The points show the estimated probability of
success, and the vertical bars the 95\% posterior credible intervals.
The orange lines show the estimates from the ordinary binary approach,
with the dashed line showing the mean and the dotted lines the 95\%
confidence interval boundaries.

We see a slight positive relationship in estimated success probability
and baseline tumour size. This is a random manifestation in the sampled
dataset. We see that the intervals are generally narrower with the
AugBin method, particularly so under the informative priors. The average
credible interval width under the informative priors is 36\% narrower
than that from the binary method. This suggests the strong effect the
priors are having.

Comparing the two sets of fitted values from the AugBin models, we see
that the estimates are generally a few percent lower under the
informtive priors. In fact, a simulation study with 1,000 iterations
reveals that the estimated probabilities of success under these
informative priors are biased downwards by 4.5\% on average, and that
the coverage of 95\% posterior credible intervals is only 71.7\%. In
contrast, simulations under the diffuse priors reveal downward bias of
approximately 1.5\% and 95\% posterior interval coverage of 92.7\%, much
closer to the nominal value. Code to run the simulation study appears in
the GitHub repository of this article at
\url{https://github.com/brockk/trialr-jss}.

As an \proglang{R}-user might expect, the \texttt{predict} function
allows the specification of \texttt{newdata} to obtain predictions for
custom cases:

\begin{CodeChunk}

\begin{CodeInput}
R> predict(fit_diffuse, newdata = data.frame(z0 = 5:10, z1 = 4:9), 
R+         y2_upper = log(0.7))
\end{CodeInput}

\begin{CodeOutput}
# A tibble: 6 x 8
     id    z0    z1 prob_success_samp prob_success lower upper ci_width
  <int> <int> <int> <list>                   <dbl> <dbl> <dbl>    <dbl>
1     1     5     4 <dbl [4,000]>            0.210 0.124 0.312    0.188
2     2     6     5 <dbl [4,000]>            0.228 0.149 0.320    0.171
3     3     7     6 <dbl [4,000]>            0.247 0.166 0.336    0.171
4     4     8     7 <dbl [4,000]>            0.265 0.177 0.362    0.185
5     5     9     8 <dbl [4,000]>            0.282 0.180 0.394    0.214
6     6    10     9 <dbl [4,000]>            0.298 0.177 0.436    0.260
\end{CodeOutput}
\end{CodeChunk}

Function parameters also allow users to specify different shrinkage
thresholds for response at interim and final timepoints.

\hypertarget{discussion}{%
\section{Discussion}\label{discussion}}

The \pkg{trialr} package exploits the power of \proglang{Stan} to offer
full Bayesian inference to clinical trialists and researchers. We have
illustrated this using the CRM and EffTox designs for conducting
dose-finding trials. One of the cornerstones of our approach is using
posterior samples for intuitive and flexible inference. For instance,
having access to posterior samples in a utility-maximising design like
EffTox allowed us to directly calculate the probabilities of superiority
of the doses. Without posterior samples, that calculation would have
been more challenging.

The other core tenets of our approach include using modern
\texttt{tidyverse} \citep{tidyverse} classes and methods and relying on
visualisation to convey insights. By using \pkg{tidybayes}
\citep{tidybayes}, posterior samples are available in tidy formats,
convenient for further analysis via \pkg{dplyr} \citep{dplyr} and
\pkg{purrr} \citep{purrr} functions, as we have demonstrated in several
examples. We have included plots using \pkg{ggplot2} \citep{ggplot2} and
graphs using \pkg{DiagrammeR} \citep{DiagrammeR}.

We have demonstrated functions and classes dedicated to calculating dose
transition pathways in dose-finding trials. Our approach will work not
just for CRM and EffTox but for general phase I and phase I/II designs.
In future releases of \pkg{trialr}, we will provide the same
funcionality for the Escalation with Overdose Control design
\citep{Tighiouart2010} and the seamless phase I/II design of
\citet{Wages2014}. Calculating dose paths for Wages \& Tait's design may
initially seem questionable because their method uses adaptive
randomisation to allocate patients to doses inferred to be tolerable. If
the dose recommended is random, what point is there in considering
future paths? Although the dose is random, the way that future
randomisation probabilities are adapted in response to outcomes will
still be of interest. Our use of \texttt{tibble} allows that a vector
may be the item of interest in a DTP analysis. For example, it would be
possible to analyse adaptive randomisation probabilities in Wages \&
Tait's design in a manner similar to how we analysed above changes in
the probability of a dose being the OBD.

We have not presented methods for conducting dose-finding simulation
studies. Though these would be simple to perform using \pkg{trialr},
users may be put off by the computational overhead of performing MCMC
sampling. For the purposes of illustration, let us assume that model
fitting by MCMC sampling takes 1 second, and that simulations would
investigate performance of a dose-finding trial using 30 patients,
evaluated in cohorts of three. Each simulated replicate would require 10
model fits, taking 10 seconds. A study of 1000 replicates would take
over 2.5 hours. By time alone, this is vastly inferior to functions that
conduct simulations using non-MCMC methods such as those offered by
\pkg{dfcrm} \citep{dfcrm}. Nevertheless, simulations remain a key tool
for clinical trialists to justify the use of such adaptive designs, and
will be developed in \pkg{trialr} in future.

We also demonstrated in this manuscript the AugBin model of
\citet{Wason2013} for single-arm trials. The authors also include a
two-arm version suitable for randomised trials. They also latterly
introduced versions of the model suitable for trials with an arbitrary
number of post-baseline assessments. These variants are not currently in
\pkg{trialr} but they will be added in the future, supporting the same
generic functions.

We investigated two sets of priors with the AugBin model. The diffuse
priors we used could hardly be described as generative or representing
any reasonable person's beliefs because they yield prior predictive
outcomes that are plainly implausible. However, our informative priors
that are broadly generative of plausible outcomes a-priori, exhibit
possibly undesirable numerical performance, with relatively large bias
and low interval coverage. Learning how to set priors for this Bayesian
version of the AugBin model is an area worthy of further research.

Despite being an \proglang{R} package, the \proglang{Stan}
implementations of the models in \pkg{trialr} can be used with
\pkg{StataStan} and \pkg{PyStan} to fit the models in \proglang{Stata}
and \proglang{Python} respectively.

\pkg{trialr} is available at
\url{https://CRAN.R-project.org/package=trialr}. Source code is
available on GitHub at \url{https://github.com/brockk/trialr}. The
source for this article is available at
\url{https://github.com/brockk/trialr-jss}.

\hypertarget{acknowledgements}{%
\section{Acknowledgements}\label{acknowledgements}}

The \pkg{trialr} package benefited from the comments of anonymous
peer-reviewers when it was presentated at StanCon, Helsinki, 2018. The
author thanks those reviewers.

\hypertarget{appendix}{%
\section{Appendix}\label{appendix}}

\hypertarget{reproducing-figures-in-crm-sections}{%
\subsection{Reproducing Figures in CRM
sections}\label{reproducing-figures-in-crm-sections}}

The following code creates Figure \ref{fig:crm_dtps_graph}. The command
\texttt{render\_graph} displays the graph in the RStudio Viewer.

\begin{CodeChunk}

\begin{CodeInput}
R> library(tibble)
R> library(dplyr)
R> paths2_df <- as_tibble(paths2)
R> 
R> library(DiagrammeR)
R> # DiagrammeR requires data.frames of nodes and edges to create a graph.
R> paths2_df %>%
R>   transmute(id = .node,
R>             type = NA,
R>             label = case_when(
R>               is.na(next_dose) ~ 'Stop',
R>               TRUE ~ next_dose %>% as.character()),
R>             shape = 'circle',
R>             fillcolor = case_when(
R>               next_dose == 1 ~ 'slategrey',
R>               next_dose == 2 ~ 'skyblue1',
R>               next_dose == 3 ~ 'royalblue1',
R>               next_dose == 4 ~ 'orchid4',
R>               next_dose == 5 ~ 'royalblue4',
R>               is.na(next_dose) ~ 'red'
R>             )
R>   ) -> nodes_df
R> paths2_df %>% 
R>   filter(!is.na(.parent)) %>% 
R>   select(from = .parent, to = .node, label = outcomes) %>% 
R>   mutate(rel = "leading_to") -> edges_df
R> graph <- create_graph(nodes_df = nodes_df, edges_df = edges_df)
R> render_graph(graph)
\end{CodeInput}
\end{CodeChunk}

\hypertarget{reproducing-figures-in-efftox-sections}{%
\subsection{Reproducing Figures in EffTox
sections}\label{reproducing-figures-in-efftox-sections}}

The following code creates Figures \ref{fig:efftox_prior_plot_1} and
\ref{fig:efftox_prior_plot_2}.

\begin{CodeChunk}

\begin{CodeInput}
R> library(tibble)
R> library(purrr)
R> library(dplyr)
R> library(tidyr)
R> library(ggplot2)
R> 
R> fit_thall_2014 <- function(outcomes, alpha_sd, beta_sd, gamma_sd, zeta_sd) {
R>   stan_efftox(outcome_str = outcomes, 
R>               real_doses = c(1.0, 2.0, 4.0, 6.6, 10.0),
R>               efficacy_hurdle = 0.5, toxicity_hurdle = 0.3,
R>               p_e = 0.1, p_t = 0.1, eff0 = 0.5, tox1 = 0.65,
R>               eff_star = 0.7, tox_star = 0.25,
R>               alpha_mean = -7.9593, alpha_sd = alpha_sd,
R>               beta_mean = 1.5482, beta_sd = beta_sd,
R>               gamma_mean = 0.7367, gamma_sd = gamma_sd,
R>               zeta_mean = 3.4181, zeta_sd = zeta_sd,
R>               eta_mean = 0, eta_sd = 0.2,
R>               psi_mean = 0, psi_sd = 1, 
R>               seed = 123, refresh = 0)
R> }
R> 
R> expand.grid(
R>   alpha_sd = c(2, 3, 4), 
R>   beta_sd = c(2, 3, 4), 
R>   gamma_sd = 2.5423,
R>   zeta_sd = 2.4406
R> ) %>% 
R>   as_tibble() %>% 
R>   mutate(fit = pmap(list(alpha_sd, beta_sd, gamma_sd, zeta_sd), 
R>                     fit_thall_2014, outcomes = ''),
R>          series = rownames(.)) -> prior_fits1
R> 
R> prior_fits1 %>% 
R>   mutate(
R>     dose = map(fit, 'dose_indices'),
R>     prob_obd = map(fit, 'prob_obd'),
R>     entropy = map_dbl(fit, 'entropy')
R>   ) %>% 
R>   select(-fit) %>% 
R>   unnest %>% 
R>   ggplot(aes(x = dose, y = prob_obd, fill = entropy)) + 
R>   geom_col() + 
R>   facet_grid(~ alpha_sd ~ beta_sd) + 
R>   labs(y = 'Prob(dose has maximal utility)', 
R>        title = 'Effect of prior SD of alpha (rows) and beta (columns)') + 
R>   theme(legend.position = 'bottom')
\end{CodeInput}
\end{CodeChunk}

\begin{CodeChunk}

\begin{CodeInput}
R> expand.grid(alpha_sd = 3.5487, 
R>             beta_sd = 3.5018,
R>             gamma_sd = c(2, 3, 4),
R>             zeta_sd = c(2, 3, 4)
R> ) %>% 
R>   as_tibble() %>% 
R>   mutate(fit = pmap(list(alpha_sd, beta_sd, gamma_sd, zeta_sd), 
R>                     fit_thall_2014, outcomes = ''),
R>          series = rownames(.)) -> prior_fits2
R> 
R> prior_fits2 %>% 
R>   mutate(
R>     dose = map(fit, 'dose_indices'),
R>     prob_obd = map(fit, 'prob_obd'),
R>     entropy = map_dbl(fit, 'entropy')
R>   ) %>% 
R>   select(-fit) %>% 
R>   unnest %>% 
R>   ggplot(aes(x = dose, y = prob_obd, fill = entropy)) + 
R>   geom_col() + 
R>   facet_grid(~ gamma_sd ~ zeta_sd) + 
R>   labs(y = 'Prob(dose has maximal utility)', 
R>        title = 'Effect of prior SD of gamma (rows) and zeta (columns)') + 
R>   theme(legend.position = 'bottom')
\end{CodeInput}
\end{CodeChunk}

The following code creates Figure \ref{fig:efftox_dtps_graph}.

\begin{CodeChunk}

\begin{CodeInput}
R> outcomes <- '1NNN 2ENN'
R> 
R> # Left-hand plot
R> paths <- efftox_dtps(cohort_sizes = c(3), previous_outcomes = outcomes, 
R>                      real_doses = c(1.0, 2.0, 4.0, 6.6, 10.0),
R>                      efficacy_hurdle = 0.5, toxicity_hurdle = 0.3,
R>                      p_e = 0.1, p_t = 0.1, eff0 = 0.5, tox1 = 0.65,
R>                      eff_star = 0.7, tox_star = 0.25,
R>                      alpha_mean = -7.9593, alpha_sd = 3.5487,
R>                      beta_mean = 1.5482, beta_sd = 3.5018,
R>                      gamma_mean = 0.7367, gamma_sd = 2.5423,
R>                      zeta_mean = 3.4181, zeta_sd = 2.4406,
R>                      eta_mean = 0, eta_sd = 0.2,
R>                      psi_mean = 0, psi_sd = 1, 
R>                      next_dose = 3, seed = 123, refresh = 0)
R> 
R> paths_df <- as_tibble(paths)
R> paths_df %>%
R>   transmute(id = .node,
R>             type = NA,
R>             label = case_when(
R>               is.na(next_dose) ~ 'Stop',
R>               TRUE ~ next_dose %>% as.character()),
R>             shape = 'circle',
R>             fillcolor = case_when(
R>               next_dose == 1 ~ 'slategrey',
R>               next_dose == 2 ~ 'skyblue1',
R>               next_dose == 3 ~ 'royalblue1',
R>               next_dose == 4 ~ 'orchid4',
R>               next_dose == 5 ~ 'royalblue4',
R>               is.na(next_dose) ~ 'red'
R>             )
R>   ) -> nodes_df
R> 
R> paths_df %>% 
R>   filter(!is.na(.parent)) %>% 
R>   select(from = .parent, to = .node, label = outcomes) %>% 
R>   mutate(rel = "leading_to") -> edges_df
R> 
R> graph <- create_graph(nodes_df = nodes_df, edges_df = edges_df)
R> render_graph(graph)
R> 
R> # Right-hand plot
R> paths <- efftox_dtps(cohort_sizes = c(1, 1), previous_outcomes = outcomes, 
R>                      next_dose = 3, real_doses = c(1.0, 2.0, 4.0, 6.6, 10.0),
R>                      efficacy_hurdle = 0.5, toxicity_hurdle = 0.3,
R>                      p_e = 0.1, p_t = 0.1, eff0 = 0.5, tox1 = 0.65,
R>                      eff_star = 0.7, tox_star = 0.25,
R>                      alpha_mean = -7.9593, alpha_sd = 3.5487,
R>                      beta_mean = 1.5482, beta_sd = 3.5018,
R>                      gamma_mean = 0.7367, gamma_sd = 2.5423,
R>                      zeta_mean = 3.4181, zeta_sd = 2.4406,
R>                      eta_mean = 0, eta_sd = 0.2,
R>                      psi_mean = 0, psi_sd = 1, 
R>                      seed = 123, refresh = 2000)
R> paths_df <- as_tibble(paths)
R> 
R> paths_df %>%
R>   transmute(id = .node,
R>             type = NA,
R>             label = case_when(
R>               is.na(next_dose) ~ 'Stop',
R>               TRUE ~ next_dose %>% as.character()),
R>             shape = 'circle',
R>             fillcolor = case_when(
R>               next_dose == 1 ~ 'slategrey',
R>               next_dose == 2 ~ 'skyblue1',
R>               next_dose == 3 ~ 'royalblue1',
R>               next_dose == 4 ~ 'orchid4',
R>               next_dose == 5 ~ 'royalblue4',
R>               is.na(next_dose) ~ 'red'
R>             )
R>   ) -> nodes_df
R> 
R> paths_df %>% 
R>   filter(!is.na(.parent)) %>% 
R>   select(from = .parent, to = .node, label = outcomes) %>% 
R>   mutate(rel = "leading_to") -> edges_df
R> 
R> graph <- create_graph(nodes_df = nodes_df, edges_df = edges_df)
R> render_graph(graph)
\end{CodeInput}
\end{CodeChunk}

\hypertarget{reproducing-figures-in-augbin-sections}{%
\subsection{Reproducing Figures in AugBin
sections}\label{reproducing-figures-in-augbin-sections}}

The following code produce Figures \ref{fig:augbin_prior_pred_tumour}.

\begin{CodeChunk}

\begin{CodeInput}
R> set.seed(123)
R> diffuse_prior_pred_data <- do.call(prior_predictive_augbin_2t_1a, 
R>                                    append(diffuse_priors, 
R>                                           list(num_samps = 1000)))
R> inf_prior_pred_data <- do.call(prior_predictive_augbin_2t_1a, 
R>                                append(informative_priors, 
R>                                       list(num_samps = 1000)))
R> 
R> library(stringr)
R> library(tidyr)
R> 
R> # Visualise tumour sizes
R> bind_rows(
R>   diffuse_prior_pred_data %>% mutate(Prior = 'Diffuse'),
R>   inf_prior_pred_data %>% mutate(Prior = 'Informative')
R> ) %>% 
R>   select(id, Prior, y0, y1, y2) %>% 
R>   gather(assessment, y, -id, -Prior) %>% 
R>   mutate(time = str_extract(assessment, '\\d+') %>% as.integer()) %>% 
R>   ggplot(aes(x = time, y = y)) + 
R>   geom_line(aes(group = id), alpha = 0.3, col = 'darkgray') + 
R>   geom_hline(yintercept = log(0.7), col = 'orange', linetype = 'dashed') + 
R>   scale_x_continuous(breaks = 0:2) + 
R>   labs(y = 'log(tumour size ratio)') + 
R>   facet_wrap(~ Prior)
\end{CodeInput}
\end{CodeChunk}

The following code produce Figures \ref{fig:augbin_prior_pred_nsf}.

\begin{CodeChunk}

\begin{CodeInput}
R> bind_rows(
R>   diffuse_prior_pred_data %>% mutate(Prior = 'Diffuse'),
R>   inf_prior_pred_data %>% mutate(Prior = 'Informative')
R> ) %>% 
R>   ggplot(aes(x = z0, y = prob_d1)) + 
R>   geom_point() + geom_smooth(method = 'gam') +
R>   labs(x = 'Baseline tumour size (cm)', 
R>        y = 'Prob(Non-shrikage failure at time 1)') + 
R>   facet_wrap(~ Prior)
\end{CodeInput}
\end{CodeChunk}

The following code produces Figure \ref{fig:post_pred_augbin}.

\begin{CodeChunk}

\begin{CodeInput}
R> pred_inf <- predict(fit_inf, y2_upper = log(0.7))
R> pred_diffuse <- predict(fit_diffuse, y2_upper = log(0.7))
R> pred_binary <- binary_prob_success(fit_diffuse, 
R>                                    y2_upper = log(0.7), 
R>                                    methods = 'exact')
R> 
R> pred_diffuse %>% 
R>   ggplot(aes(x = z0, y = prob_success, col = z1)) + 
R>   geom_point() + 
R>   geom_linerange(aes(ymin = lower, ymax = upper)) + 
R>   geom_hline(yintercept = pred_binary$mean, col = 'orange', 
R>              linetype = 'dashed') +
R>   geom_hline(yintercept = pred_binary$lower, col = 'orange', 
R>              linetype = 'dotted') +
R>   geom_hline(yintercept = pred_binary$upper, col = 'orange', 
R>              linetype = 'dotted') +
R>   ylim(0, 0.75) + 
R>   labs(x = 'Baseline tumour size (cm)', y = 'Prob(Success)', 
R>        col = 'Interim size (cm)', title = 'Diffuse') + 
R>   theme(legend.position = 'bottom') -> aubgin_diffuse_plot
R> 
R> pred_inf %>% 
R>   ggplot(aes(x = z0, y = prob_success, col = z1)) + 
R>   geom_point() + 
R>   geom_linerange(aes(ymin = lower, ymax = upper)) + 
R>   geom_hline(yintercept = pred_binary$mean, col = 'orange', 
R>              linetype = 'dashed') +
R>   geom_hline(yintercept = pred_binary$lower, col = 'orange', 
R>              linetype = 'dotted') +
R>   geom_hline(yintercept = pred_binary$upper, col = 'orange', 
R>              linetype = 'dotted') +
R>   ylim(0, 0.75) + 
R>   labs(x = 'Baseline tumour size (cm)', y = 'Prob(Success)', 
R>        col = 'Interim size (cm)', title = 'Informative') + 
R>   theme(legend.position = 'bottom') -> aubgin_inf_plot
R> 
R> gridExtra::grid.arrange(aubgin_diffuse_plot, aubgin_inf_plot, ncol = 2)
\end{CodeInput}
\end{CodeChunk}

\bibliography{bibliography.bib,extrabib.bib}

\begin{thebibliography}{37}
\newcommand{\enquote}[1]{``#1''}
\providecommand{\natexlab}[1]{#1}
\providecommand{\url}[1]{\texttt{#1}}
\providecommand{\urlprefix}{URL }
\expandafter\ifx\csname urlstyle\endcsname\relax
  \providecommand{\doi}[1]{doi:\discretionary{}{}{}#1}\else
  \providecommand{\doi}{doi:\discretionary{}{}{}\begingroup
  \urlstyle{rm}\Url}\fi
\providecommand{\eprint}[2][]{\url{#2}}

\bibitem[{Braun(2002)}]{Braun2002}
Braun TM (2002).
\newblock \enquote{The Bivariate Continual Reassessment Method: {{Extending}}
  the {{CRM}} to Phase {{I}} Trials of Two Competing Outcomes.}
\newblock \emph{Controlled Clinical Trials}, \textbf{23}(3), 240--256.
\newblock ISSN 01972456.
\newblock \doi{10.1016/S0197-2456(01)00205-7}.

\bibitem[{Brock(2019)}]{brockMethodsIncreaseEfficiency}
Brock K (2019).
\newblock \emph{Methods to Increase Efficiency in Clinical Trials with
  Restricted Sample Size}.
\newblock Ph.D. thesis, University of Birmingham.
\newblock
  \urlprefix\url{https://etheses.bham.ac.uk/id/eprint/8921/1/Brock2019PhD.pdf}.

\bibitem[{Brock \emph{et~al.}(2017)Brock, Billingham, Copland, Siddique,
  Sirovica, and Yap}]{Brock2017a}
Brock K, Billingham L, Copland M, Siddique S, Sirovica M, Yap C (2017).
\newblock \enquote{Implementing the {{EffTox}} Dose-Finding Design in the
  {{Matchpoint}} Trial.}
\newblock \emph{BMC Medical Research Methodology}, \textbf{17}(1), 112.
\newblock ISSN 1471-2288.
\newblock \doi{10.1186/s12874-017-0381-x}.

\bibitem[{Carpenter \emph{et~al.}(2016)Carpenter, Gelman, Hoffman, Lee,
  Goodrich, Betancourt, Brubaker, Li, and Riddell}]{Carpenter2016}
Carpenter B, Gelman A, Hoffman M, Lee D, Goodrich B, Betancourt M, Brubaker MA,
  Li P, Riddell A (2016).
\newblock \enquote{Stan: {{A Probabilistic Programming Language}}.}
\newblock \emph{Journal of Statistical Software}, \textbf{76}(Ii), 1--32.

\bibitem[{Carter(1973)}]{carterStudyDesignPrinciples1973}
Carter S (1973).
\newblock \enquote{Study Design Principles for the Clinical Evaluation of New
  Drugs as Developed by the Chemotherapy Programme of the {{National Cancer
  Institute}}.}
\newblock In \emph{The {{Design}} of {{Clinical Trials}} in {{Cancer
  Therapy}}}, pp. 242--289. {Editions Scientifique Europe}.

\bibitem[{Cheung(2013)}]{dfcrm}
Cheung K (2013).
\newblock \emph{dfcrm: Dose-finding by the continual reassessment method}.
\newblock R package version 0.2-2,
  \urlprefix\url{https://CRAN.R-project.org/package=dfcrm}.

\bibitem[{Cheung(2011)}]{Cheung2011}
Cheung YK (2011).
\newblock \emph{Dose {{Finding}} by the {{Continual Reassessment Method}}}.
\newblock {Chapman \& Hall / CRC Press}, {New York}.
\newblock ISBN 978-1-4200-9151-9.

\bibitem[{Cheung and Chappell(2000)}]{Cheung2000}
Cheung YK, Chappell R (2000).
\newblock \enquote{Sequential Designs for Phase {{I}} Clinical Trials with
  Late-Onset Toxicities.}
\newblock \emph{Biometrics}, \textbf{56}(4), 1177--1182.
\newblock ISSN 0006-341X.

\bibitem[{Chiuzan \emph{et~al.}(2017)Chiuzan, Shtaynberger, Manji, Duong,
  Schwartz, Ivanova, and Lee}]{Chiuzan2017}
Chiuzan C, Shtaynberger J, Manji GA, Duong JK, Schwartz GK, Ivanova A, Lee SM
  (2017).
\newblock \enquote{Dose-Finding Designs for Trials of Molecularly Targeted
  Agents and Immunotherapies.}
\newblock \emph{Journal of Biopharmaceutical Statistics}, \textbf{27}(3),
  477--494.
\newblock ISSN 1054-3406.
\newblock \doi{10.1080/10543406.2017.1289952}.

\bibitem[{Craddock \emph{et~al.}(2019)Craddock, Slade, De~Santo, Wheat,
  Ferguson, Hodgkinson, Brock, Cavenagh, Ingram, Dennis, Malladi, Siddique,
  Mussai, and Yap}]{craddockCombinationLenalidomideAzacitidine2019}
Craddock C, Slade D, De~Santo C, Wheat R, Ferguson P, Hodgkinson A, Brock K,
  Cavenagh J, Ingram W, Dennis M, Malladi R, Siddique S, Mussai F, Yap C
  (2019).
\newblock \enquote{Combination {{Lenalidomide}} and {{Azacitidine}}: {{A Novel
  Salvage Therapy}} in {{Patients Who Relapse After Allogeneic Stem}}-{{Cell
  Transplantation}} for {{Acute Myeloid Leukemia}}.}
\newblock \emph{Journal of Clinical Oncology}, p. JCO.18.00889.
\newblock ISSN 0732-183X, 1527-7755.
\newblock \doi{10.1200/JCO.18.00889}.

\bibitem[{Dorai-Raj(2014)}]{binom}
Dorai-Raj S (2014).
\newblock \emph{binom: Binomial Confidence Intervals For Several
  Parameterizations}.
\newblock R package version 1.1-1,
  \urlprefix\url{https://CRAN.R-project.org/package=binom}.

\bibitem[{Eisenhauer \emph{et~al.}(2009)Eisenhauer, Therasse, Bogaerts,
  Schwartz, Sargent, Ford, Dancey, Arbuck, Gwyther, Mooney, Rubinstein,
  Shankar, Dodd, Kaplan, Lacombe, and Verweij}]{Eisenhauer2009}
Eisenhauer EA, Therasse P, Bogaerts J, Schwartz LH, Sargent D, Ford R, Dancey
  J, Arbuck S, Gwyther S, Mooney M, Rubinstein L, Shankar L, Dodd L, Kaplan R,
  Lacombe D, Verweij J (2009).
\newblock \enquote{New Response Evaluation Criteria in Solid Tumours: {{Revised
  RECIST}} Guideline (Version 1.1).}
\newblock \emph{European Journal of Cancer}, \textbf{45}(2), 228--247.
\newblock ISSN 09598049.
\newblock \doi{10.1016/j.ejca.2008.10.026}.

\bibitem[{Henry and Wickham(2019)}]{purrr}
Henry L, Wickham H (2019).
\newblock \emph{purrr: Functional Programming Tools}.
\newblock R package version 0.3.2,
  \urlprefix\url{https://CRAN.R-project.org/package=purrr}.

\bibitem[{Herbst \emph{et~al.}(2016)Herbst, Baas, Kim, Felip,
  {P{\'e}rez-Gracia}, Han, Molina, Kim, Arvis, Ahn, Majem, Fidler, De~Castro,
  Garrido, Lubiniecki, Shentu, Im, {Dolled-Filhart}, and Garon}]{Herbst2016}
Herbst RS, Baas P, Kim DW, Felip E, {P{\'e}rez-Gracia} JL, Han JY, Molina J,
  Kim JH, Arvis CD, Ahn MJ, Majem M, Fidler MJ, De~Castro G, Garrido M,
  Lubiniecki GM, Shentu Y, Im E, {Dolled-Filhart} M, Garon EB (2016).
\newblock \enquote{Pembrolizumab versus Docetaxel for Previously Treated,
  {{PD}}-{{L1}}-Positive, Advanced Non-Small-Cell Lung Cancer
  ({{KEYNOTE}}-010): {{A}} Randomised Controlled Trial.}
\newblock \emph{The Lancet}, \textbf{387}(10027), 1540--1550.
\newblock ISSN 1474547X.
\newblock \doi{10.1016/S0140-6736(15)01281-7}.

\bibitem[{Iannone(2018)}]{DiagrammeR}
Iannone R (2018).
\newblock \emph{DiagrammeR: Graph/Network Visualization}.
\newblock R package version 1.0.0,
  \urlprefix\url{https://CRAN.R-project.org/package=DiagrammeR}.

\bibitem[{Iasonos \emph{et~al.}(2008)Iasonos, Wilton, Riedel, Seshan, and
  Spriggs}]{Iasonos2008}
Iasonos A, Wilton AS, Riedel ER, Seshan VE, Spriggs DR (2008).
\newblock \enquote{A Comprehensive Comparison of the Continual Reassessment
  Method to the Standard 3 + 3 Dose Escalation Scheme in {{Phase I}}
  Dose-Finding Studies.}
\newblock \emph{Clinical trials (London, England)}, \textbf{5}(5), 465--477.
\newblock ISSN 1740-7745.
\newblock \doi{10.1177/1740774508096474}.

\bibitem[{Ji and Wang(2013)}]{Ji2013a}
Ji Y, Wang SJ (2013).
\newblock \enquote{Modified Toxicity Probability Interval Design: A Safer and
  More Reliable Method than the 3 + 3 Design for Practical Phase {{I}} Trials.}
\newblock \emph{Journal of clinical oncology : official journal of the American
  Society of Clinical Oncology}, \textbf{31}(14), 1785--1791.
\newblock ISSN 15277755.
\newblock \doi{10.1200/JCO.2012.45.7903}.

\bibitem[{Kay(2019)}]{tidybayes}
Kay M (2019).
\newblock \emph{{tidybayes}: Tidy Data and Geoms for {Bayesian} Models}.
\newblock \doi{10.5281/zenodo.1308151}.
\newblock R package version 1.0.4,
  \urlprefix\url{http://mjskay.github.io/tidybayes/}.

\bibitem[{Le~Tourneau \emph{et~al.}(2009)Le~Tourneau, Lee, and
  Siu}]{LeTourneau2009}
Le~Tourneau C, Lee JJ, Siu LL (2009).
\newblock \enquote{Dose Escalation Methods in Phase i Cancer Clinical Trials.}
\newblock \emph{Journal of the National Cancer Institute}, \textbf{101}(10),
  708--720.
\newblock ISSN 00278874.
\newblock \doi{10.1093/jnci/djp079}.

\bibitem[{O'Quigley and
  Paoletti(2003)}]{oquigleyContinualReassessmentMethod2003}
O'Quigley J, Paoletti X (2003).
\newblock \enquote{Continual {{Reassessment Method}} for {{Ordered Groups}}.}
\newblock \emph{Biometrics}, \textbf{59}(2), 430--440.
\newblock ISSN 1541-0420.
\newblock \doi{10.1111/1541-0420.00050}.

\bibitem[{O'Quigley \emph{et~al.}(1990)O'Quigley, Pepe, and
  Fisher}]{OQuigley1990}
O'Quigley J, Pepe M, Fisher L (1990).
\newblock \enquote{Continual Reassessment Method: A Practical Design for Phase
  1 Clinical Trials in Cancer.}
\newblock \emph{Biometrics}, \textbf{46}(1), 33--48.
\newblock ISSN 0006-341X.
\newblock \doi{10.2307/2531628}.

\bibitem[{O'Quigley and Zohar(2006)}]{oquigleyExperimentalDesignsPhase2006}
O'Quigley J, Zohar S (2006).
\newblock \enquote{Experimental Designs for Phase {{I}} and Phase {{I}}/{{II}}
  Dose-Finding Studies.}
\newblock \emph{British Journal of Cancer}, \textbf{94}(5), 609--613.
\newblock ISSN 0007-0920.
\newblock \doi{10.1038/sj.bjc.6602969}.

\bibitem[{Rogatko \emph{et~al.}(2007)Rogatko, Schoeneck, Jonas, Tighiouart,
  Khuri, and Porter}]{Rogatko2007}
Rogatko A, Schoeneck D, Jonas W, Tighiouart M, Khuri FR, Porter A (2007).
\newblock \enquote{Translation of Innovative Designs into Phase {{I}} Trials.}
\newblock \emph{Journal of Clinical Oncology}, \textbf{25}(31), 4982--4986.
\newblock ISSN 0732183X.
\newblock \doi{10.1200/JCO.2007.12.1012}.

\bibitem[{{Sabanes Bove} \emph{et~al.}(2018){Sabanes Bove}, {Yin Yeung},
  Palermo, and Jaki}]{crmPack}
{Sabanes Bove} D, {Yin Yeung} W, Palermo G, Jaki T (2018).
\newblock \emph{crmPack: Object-Oriented Implementation of CRM Designs}.
\newblock R package version 0.2.9,
  \urlprefix\url{https://CRAN.R-project.org/package=crmPack}.

\bibitem[{Sweeting \emph{et~al.}(2013)Sweeting, Mander, and Sabin}]{bcrm}
Sweeting M, Mander A, Sabin T (2013).
\newblock \enquote{{bcrm}: Bayesian Continual Reassessment Method Designs for
  Phase I Dose-Finding Trials.}
\newblock \emph{Journal of Statistical Software}, \textbf{54}(13), 1--26.
\newblock \urlprefix\url{http://www.jstatsoft.org/article/view/v054i13}.

\bibitem[{Thall and Cook(2004)}]{Thall2004}
Thall P, Cook J (2004).
\newblock \enquote{Dose-{{Finding Based}} on {{Efficacy}}-{{Toxicity
  Trade}}-{{Offs}}.}
\newblock \emph{Biometrics}, \textbf{60}(3), 684--693.

\bibitem[{Thall \emph{et~al.}(2006)Thall, Cook, and Estey}]{Thall2006}
Thall P, Cook J, Estey E (2006).
\newblock \enquote{Adaptive Dose Selection Using Efficacy-Toxicity Trade-Offs:
  Illustrations and Practical Considerations.}
\newblock \emph{Journal of biopharmaceutical statistics}, \textbf{16}(5),
  623--638.
\newblock ISSN 1054-3406.
\newblock \doi{10.1080/10543400600860394}.

\bibitem[{Thall \emph{et~al.}(2014)Thall, Herrick, Nguyen, Venier, and
  Norris}]{Thall2014}
Thall P, Herrick R, Nguyen H, Venier J, Norris J (2014).
\newblock \enquote{Effective Sample Size for Computing Prior Hyperparameters in
  {{Bayesian}} Phase {{I}}-{{II}} Dose-Finding.}
\newblock \emph{Clinical Trials}, \textbf{11}(6), 657--666.
\newblock ISSN 1740-7745.
\newblock \doi{10.1177/1740774514547397}.

\bibitem[{Thall \emph{et~al.}(2003)Thall, Wathen, Bekele, Champlin, Baker, and
  Benjamin}]{Thall2003}
Thall PF, Wathen JK, Bekele BN, Champlin RE, Baker LH, Benjamin RS (2003).
\newblock \enquote{Hierarchical {{Bayesian}} Approaches to Phase {{II}} Trials
  in Diseases with Multiple Subtypes.}
\newblock \emph{Statistics in Medicine}, \textbf{22}(5), 763--780.
\newblock ISSN 02776715.
\newblock \doi{10.1002/sim.1399}.

\bibitem[{Tighiouart and Rogatko(2010)}]{Tighiouart2010}
Tighiouart M, Rogatko A (2010).
\newblock \enquote{Dose {{Finding}} with {{Escalation}} with {{Overdose
  Control}} ({{EWOC}}) in {{Cancer Clinical Trials}}.}
\newblock \emph{Statistical Science}, \textbf{25}(2), 217--226.
\newblock ISSN 0883-4237.
\newblock \doi{10.1214/10-STS333}.
\newblock \eprint{1011.6479v1}.

\bibitem[{Wages and Tait(2015)}]{Wages2014}
Wages NA, Tait C (2015).
\newblock \enquote{Seamless {{Phase I}}/{{II Adaptive Design For Oncology
  Trials}} of {{Molecularly Targeted Agents}}.}
\newblock \emph{Journal of biopharmaceutical statistics}, pp. 1--30.
\newblock ISSN 1520-5711.
\newblock \doi{10.1080/10543406.2014.920873}.

\bibitem[{Wason and Seaman(2013)}]{Wason2013}
Wason JMS, Seaman SR (2013).
\newblock \enquote{Using Continuous Data on Tumour Measurements to Improve
  Inference in Phase {{II}} Cancer Studies.}
\newblock \emph{Statistics in Medicine}, \textbf{32}(26), 4639--4650.
\newblock ISSN 02776715.
\newblock \doi{10.1002/sim.5867}.

\bibitem[{Wickham(2016)}]{ggplot2}
Wickham H (2016).
\newblock \emph{ggplot2: Elegant Graphics for Data Analysis}.
\newblock Springer-Verlag New York.
\newblock ISBN 978-3-319-24277-4.
\newblock \urlprefix\url{https://ggplot2.tidyverse.org}.

\bibitem[{Wickham(2017)}]{tidyverse}
Wickham H (2017).
\newblock \emph{tidyverse: Easily Install and Load the 'Tidyverse'}.
\newblock R package version 1.2.1,
  \urlprefix\url{https://CRAN.R-project.org/package=tidyverse}.

\bibitem[{Wickham \emph{et~al.}(2019)Wickham, François, Henry, and
  Müller}]{dplyr}
Wickham H, François R, Henry L, Müller K (2019).
\newblock \emph{dplyr: A Grammar of Data Manipulation}.
\newblock R package version 0.8.1,
  \urlprefix\url{https://CRAN.R-project.org/package=dplyr}.

\bibitem[{Yap \emph{et~al.}(2017)Yap, Billingham, Cheung, Craddock, and
  O'Quigley}]{Yap2017}
Yap C, Billingham LJ, Cheung YK, Craddock C, O'Quigley J (2017).
\newblock \enquote{Dose Transition Pathways: {{The}} Missing Link between
  Complex Dose-Finding Designs and Simple Decision-Making.}
\newblock \emph{Clinical Cancer Research}, \textbf{23}(24), 7440--7447.
\newblock ISSN 15573265.
\newblock \doi{10.1158/1078-0432.CCR-17-0582}.

\bibitem[{Zhang \emph{et~al.}(2006)Zhang, Sargent, and Mandrekar}]{Zhang2006}
Zhang W, Sargent DJ, Mandrekar S (2006).
\newblock \enquote{An Adaptive Dose-Finding Design Incorporating Both Toxicity
  and Efficacy.}
\newblock \emph{Statistics in Medicine}, \textbf{25}(14), 2365--2383.
\newblock ISSN 02776715.
\newblock \doi{10.1002/sim.2325}.

\end{thebibliography}

\end{document}